\documentstyle[12pt]{article}
\begin{document}
\def\lsim{\:\raisebox{-0.5ex}{$\stackrel{\textstyle<}{\sim}$}\:}
\def\gsim{\:\raisebox{-0.5ex}{$\stackrel{\textstyle>}{\sim}$}\:}
\begin{flushright}
TIFR/TH/93-40 \\
\end{flushright}
\bigskip

\begin{center}
{{\bf ELECTROWEAK PROCESSES AND PRECISION TESTS}\footnote{Lectures
given at the {\it Winter School on Particle Physics and Cosmology at the
Interface} (2--17 January, 1993, Puri, India), to appear in the
Proceedings to be published by {\it World Scientific.}}}

\vspace {1.5cm}

Probir Roy\\
Tata Institute of Fundamental Research\\
Homi Bhabha Road, Bombay 400 005\\
India

\end{center}

\vspace {1.5 cm}

\begin{center}
CONTENTS
\end{center}

\vspace{.5cm}

\begin{itemize}

\item Review of the electroweak sector of the \\
minimal standard model

\item Basic LEP processes at the tree level \\

\item 1-loop radiative corrections in the on-shell \\
renormalization scheme

\item Star scheme and oblique corrections

\item Introduction to oblique parameters

\end{itemize}

\newpage

\begin{itemize}

\item {\bf Review of the electroweak sector of the Minimal Standard Model}

\end{itemize}

The electroweak theory of Glashow-Weinberg-Salam [1] and the quantum
chromodynamic [2] description of strong interactions together comprise the
Standard Model (SM) of particle physics.  The interactions of this theory
are derived by use of the local gauge principle based on the gauge group
structure

\vspace {3 cm}
\begin{center}
Fig. 1. Gauge group structure of SM
\end{center}

\noindent The Higgs mechanism [3] screens out a part of the electroweak
forces which then constitutes the short-range weak interactions.  Left
residually is an exact abelian long-range quantum electrodynamic
interaction characterized by a conserved electromagnetic charge $Q^{EM}$.
The Higgs mechanism does not touch the QCD part which remains an exact
confining $SU(3)$ gauge theory.  We shall not deal with this sector,
leaving it to the lectures of D.P. Roy.

Let us first remark on the behavior of three generations of matter
particles/fields with respect to these gauge groups.  $SU(3)_C$ recognizes
the fundamental color triplet of quarks for each of six flavors and treats
left- and right-chiral ones on equal footing leaving all neutrinos and
charged leptons as color singlets.  $SU(2)_L$ distinguishes between flavor
doublets of left-chiral fermions $f_L = \frac{1}{2} (1 - \gamma_5)f$ with
$T_{3L} = +\frac{1}{2}$ (up type), $-\frac{1}{2}$ (down type) and singlets
of right-chiral fermions $f_R = \frac{1}{2} (1 + \gamma_5)f$ with $T_{3R}
\equiv 0$.  Of the twelve different fermions, eleven have values or upper
limits on their masses [4] but the twelfth, namely the top quark, only has
a lower limit so far on its mass -- in the vicinity of 113 GeV.  The
remaining factor group $U(1)_Y$ attributes a weak hypercharge $Y$ to each
chiral fermion which is, in general, different for the $L$- and
$R$-components.  $Y$ takes values given by the weak Gell-Mann-Nishijima formula
$$
Q = T_{3L} + \frac{1}{2} Y,
$$
with $Q_\nu = 0$, $Q_{e,\mu,\tau} = -1$, $Q_{u,c,t} = \frac{2}{3}$,
$Q_{d,s,b} = -\frac{1}{3}$ in units of the positron charge.  The
corresponding properties of antiparticles can be simply obtained by
$C$-conjugation.
$$
\begin{array}{l}
\left(\matrix{\nu_e \cr e}\right)_L, ~\left(\matrix{u \cr d}\right)_L;
{}~e_R, u_R, d_R. \\[4mm]
\left(\matrix{\nu_\mu \cr \mu}\right)_L, ~\left(\matrix{c \cr
s}\right)_L;~ \mu_R, c_R , s_R. \\[4mm]
\left(\matrix{\nu_\tau \cr \tau}\right)_L, ~\left(\matrix{t \cr
b}\right)_L; ~t_R,b_R,\tau_R.
\end{array}
$$
\begin{center}
Fig. 2. Fermion content of SM.
\end{center}

Local gauge invariance -- in the absence of any spontaneous symmetry
breakdown -- makes the following requirement.  To every generator of each
factor group of $G_{SM}$, there must correspond a massless gauge boson
coupling minimally to matter fields;  furthermore, all gauge bosons
in a simple factor group must have a universal coupling strength.  In
the electroweak sector they are $W^a_\mu ~(a = 1,2,3)$ and $B_\mu$ with
$SU(2)$ and $U(1)$ gauge coupling strengths $g$ and $g'$ respectively.
The fermion-gauge interactions are then included and specified in the
generalized fermion kinetic energy terms
$$
{\cal L}_{fg} = i \sum_{f,a} \left[\bar f_L \gamma^\mu \left(\partial_\mu
- igW^a_\mu T^a - ig' B_\mu \frac{Y}{2}\right) f_L + \bar f_R \gamma^\mu
\left(\partial_\mu - ig' B_\mu \frac{Y}{2}\right)f_R\right]
\eqno (1)
$$
in the Lagrangian density.  In (1) $f$ is a generic fermion, $a$ sums over 1
to 3 and $T^a = \frac{1}{2} \tau^a$.  The generalized gauge boson kinetic
energy terms are
$$
{\cal L}_{gb} = -\frac{1}{4} (\partial_\mu B_\nu - \partial_\nu B_\mu)^2 -
\frac{1}{4} (\partial_\mu W^a_\nu - \partial_\nu W^a_\mu + ig
\epsilon^{abc} W^b_\mu W^c_\nu)^2.
\eqno (2)
$$

The Higgs mechanism generates masses for the weak bosons via the
spontaneous breakdown of $SU(2)_L \times U(1)_Y$ to $U(1)_{EM}$.  ~This ~is
{}~done ~in

\newpage

\noindent the Minimal Standard Model (MSM) in a certain way and that is
perhaps where the physics that lies beyond is most likely to show up.  The
procedure is to introduce a complex doublet Higgs scalar field $\phi$ with
$Y = 1$:
$$
\phi = \left({\phi^+ \atop \phi^0}\right), ~~~\phi^c = \left({\bar\phi^0
\atop -\phi^-}\right).
$$
This leads to a gauge-Higgs term as a generalized Higgs kinetic energy:
$$
{\cal L}_{gh} = \sum_a \Bigg|\left(\partial_\mu - igW^a_\mu T^a - ig'
B_\mu \frac{Y}{2}\right)\phi\Bigg|^2.
\eqno (3)
$$
Furthermore, gauge invariant and renormalizable quadratic and quartic
self-Higgs terms
$$
{\cal L}_{hh} = -V(\phi) \equiv \mu^2 |\phi|^2 - \lambda |\phi|^4,
\eqno (4)
$$
with $\lambda,\mu^2 > 0$ are postulated.  The minimization of $V(\phi)$
makes $\phi$ acquire a vacuum expectation value (VEV) which
is real (since any phase can be rotated away):
$$
\langle \phi \rangle = \left({0 \atop v/\sqrt{2}}\right), ~~~\langle
\phi^c \rangle = \left({v/\sqrt{2} \atop 0}\right).
\eqno (5)
$$

The introduction of the VEV (5) into (3) leads to the gauge boson mass terms
$$
{\cal L}_M = -\frac{1}{2} \left[\frac{1}{2} (g v)\right]^2
\left[(W^1_\mu)^2 + (W^2_\mu)^2\right] - \frac{1}{2} v^2
\left[\frac{1}{2} (gW^3_\mu - g'B_\mu)\right]^2.
\eqno (6)
$$
(6) can be rewritten as
$$
-\frac{1}{2} M^2_Z Z_\mu Z^\mu - \frac{1}{2} M^2_W W^+_\mu W^{\mu -},
$$
provided we identify the gauge boson mass eigenstates and eigenvalues as
$$
\begin{array}{l}
W^\pm_\mu = \frac{1}{\sqrt{2}} \left(W^1_\mu \mp iW^2_\mu\right), ~~M_W =
\frac{1}{2} gv.  \\[2mm]
Z_\mu = (g^2 + g^{\prime^2})^{-1/2} (gW^3_\mu - g' B_\mu), ~~M_Z =
\frac{1}{2} (g^2 + g^{\prime^2})^{1/2} v.  \\[2mm]
A_\mu = (g^2 + g^{\prime^2})^{-1/2} (g' W^3_\mu + g B_\mu), ~~M_\gamma =
0.
\end{array}
\eqno (7)
$$
Let us introduce mixing parameters $c_\theta$, $s_\theta$, with
$c^2_\theta + s^2_\theta = 1$, such that
$$
c_\theta = g(g^2 + g^{\prime^2})^{-1/2}, ~~~s_\theta = g' (g^2 +
g^{\prime^2})^{-1/2},
$$
i.e. $t_\theta \equiv s_\theta/c_\theta = g'/g$ and $Z_\mu = c_\theta
W^3_\mu - s_\theta B_\mu$ while $A_\mu = s_\theta W^3_\mu + c_\theta
B_\mu$.  It follows that
$$
\frac{M^2_W}{M^2_Z c^2_\theta} = 1.
\eqno (8)
$$
(8), in fact, turns out to be true (at the tree level) not only in the MSM
but also in any $SU(2)_L \times U(1)_Y$ model with arbitrary elementary
Higgs fields so long as only the $SU(2)_L$ doublets among them have
neutral components acquiring VEVs.  Indeed, it is valid even in condensate
models without elementary scalar fields (e.g. technicolor [5]) provided
there is a global custodial isospin [6] invariance protecting the
symmetry-breaking sector.  (8) is experimentally known to be quite
accurate and we shall {\it always} assume it at the tree level.  However,
loop corrections make the LHS of (8) deviate from unity and then it is
called the $\rho$-parameter, to be defined more precisely later.

(1) can now be rewritten, with $f$ being a generic fermion field and $e =
gs_\theta$, as
\begin{eqnarray}
{\cal L}_{fg} &=& i \sum_f \bar f \partial\!\!\!/ f + g/\sqrt{2}
(J^+_{\mu L} W^{\mu^-} + h.c.) + \displaystyle {g \over c_\theta}
J_\mu^{NC} Z^\mu + e J^Q_\mu A^\mu \nonumber\\ [2mm]
&=& \sum_f i(f_L \partial\!\!\!/ f_L + \bar f_R \partial\!\!\!/ f_R) +
\displaystyle {e \over \sqrt{2} s_\theta} (J^+_{\mu L} W^{\mu^-} + h.c.)
\nonumber \\[2mm]
& & ~~~~~~~~~~~~~~~+
\displaystyle {e \over s_\theta c_\theta} J^{NC}_\mu Z^\mu + e J^Q_\mu
A^\mu. \nonumber\\[2mm]
& &
{}~~~~~~~~~~~~~~~~~~~~~~~~~~~~~~~~~~~~~~~~~~~~~~~~~~~~~~~~~~~
(9)
\nonumber
\end{eqnarray}
In (9) the weak charged, weak neutral and electromagnetic currents are
respectively given, with $J^-_{\mu L} = (J^+_{\mu L})^\dagger$, by
\begin{eqnarray}
J^+_{\mu L} = \sum_f \bar f_L \gamma_\mu T^+ f_L &=& \sum_\ell \bar
\nu_{eL} \gamma_\mu \ell_L + (\bar u_L  \bar c_L \bar
t_L) \gamma_\mu V_{CKM} \left(\matrix{d_L \cr s_L \cr b_L}\right),
\nonumber\\[2mm]
&=& \sum_\ell \bar \nu_L \frac{1}{2} \gamma_\mu (1 - \gamma_5)\ell + (\bar
u \ \bar c \ \bar t) \frac{1}{2} \gamma_\mu (1 - \gamma_5)
V_{CKM} \left(\matrix{d \cr s \cr b}\right),\nonumber  \\[2mm]
& & ~~~~~~~~~~~~~~~~~~~~~~~~~~~~~~~~~~~~~~~~~~~~~~~~~~~~~~~~~ (10) \nonumber
\end{eqnarray}
\begin{eqnarray}
J^{NC}_\mu = J^3_{\mu L} - s^2_\theta J^Q_\mu &=& \sum_f (\bar f_L
\gamma_\mu T_3 f_L - s^2_\theta Q_f \bar f \gamma_\mu f) \nonumber\\[2mm]
&=& \sum_f \bar f \frac{1}{2} \gamma_\mu \left[(T_3 - 2s^2_\theta Q_f) -
T_3 \gamma_5\right] f, ~~~~~~~~~~~~~~~ (11) \nonumber
\end{eqnarray}
$$
J^Q_\mu = \sum_f Q_f \bar f \gamma_\mu f = \sum_f Q_f (\bar f_L \gamma_\mu
f_L + \bar f_R \gamma_\mu f_R).
\eqno (12)
$$
In (10) $\ell$ sums over the charged leptons $e,\mu,\tau$ while in (11)
and (12) $Q_f$ is the electromagnetic charge of $f$.  $V_{CKM} =
(V^\dagger_{CKM})^{-1}$ is the unitary Cabibbo-Kobayashi-Maskawa $3 \times
3$ flavor matrix [7]:
$$
\pmatrix{V_{ud} & V_{us} & V_{ub} \cr V_{cd} & V_{cs} & V_{cb} \cr V_{td}
& V_{ts} & V_{tb}}.
$$
In the charged current $J^\pm_{\mu L} = J^1_{\mu L} \pm iJ^2_{\mu L}$ the
quarks can change flavor through $V_{CKM}$ while the neutral current is
strictly flavor-conserving at the tree level by virtue of the GIM
mechanism.  On the other
hand, the charged current is purely left-chiral whereas the neutral
current has the chirally asymmetric combination $T_{3L} \gamma_\mu (1 -
\gamma_5) - 2s^2_\theta Q_f$.  The latter is a vivid demonstration of the
unification between the $(V-A)$ weak and the $(Q)$ electromagnetic
charges.  We shall rewrite (12) as
$$
J^{NC}_\mu = \frac{1}{2} \sum_f \bar f \gamma_\mu (v_f - a_f \gamma_5)f,
\eqno (13)
$$
with
$$
v_f = T_3 - 2Qs^2_\theta, ~~~a_f = T_3.
$$
Thus, for instance, $v_e = -\frac{1}{2} (1 - 4s^2_\theta)$ and $a_e = -1/2$.

At high energies (such as at LEP 1 with $\sqrt{s} \simeq 90$ GeV), the
interactions among fermions and gauge bosons can be explicitly studied
through distinct signatures of the $W,Z$ bosons.  At an energy or four
momentum transfer sq. root $\sqrt{q^2}$ much below $M_W,M_Z$, however,
only the effective four-fermion interaction, mediated by $W,Z$-exchange in
a four-legged tree diagram (such as that for mu-decay, Fig. 3a or $\nu_\mu
e$ elastic scattering, Fig. 3b) is available for experimental study.

\vspace* {3.5 cm}
\begin{center}
Fig. 3. Tree-level four-fermion weak processes
\end{center}

\noindent The effective interaction in the pointlike limit becomes
$$
{\cal L}_{eff} = \frac{4G_F}{\sqrt{2}} \left[J^+_{\mu L} J^{\mu^-}_L +
J^{NC}_\mu J^{\mu^{NC}}\right],
\eqno (14)
$$
with the Fermi constant given by
$$
\frac{G_F}{\sqrt{2}} = \frac{g^2}{8M^2_W} =
\frac{4\pi\alpha_{EM}}{8s^2_\theta c^2_\theta M^2_Z},
\eqno (15)
$$
where $\alpha_{EM}$ is the fine structure constant.  This means, of
course, that $v = (\sqrt{2} G_F)^{-1/2} \simeq 246$ GeV.  Low energy weak
scattering and decay data are now known to conform to the MSM at an
accuracy $\lsim 5\%$.

Fermion mass terms arise as a consequence of the spontaneous symmetry
breakdown from Yukawa interactions among $f,\bar f$ and $\phi$.  They can
be written as
$$
- \sum_f m_f \bar f f.
$$
The shift $\phi \rightarrow \phi' \equiv \phi - \langle \phi \rangle$ from
unphysical fields to physical normal modes does two things.  (1) It
transforms the components $\phi^\pm$ and $\chi \equiv (\sqrt{2} i)^{-1}
(\phi^0 - \bar \phi^0)$ into longitudinal (Goldstone) $W^\pm_L$ and $Z_L$
components respectively; (2) it leaves residually in the spectrum a
single Higgs field $H = (\sqrt{2})^{-1} (\phi^0 + \bar \phi^0 - \sqrt{2}
v)$ corresponding to a scalar particle of mass $m_H = (\sqrt{2}
\lambda)^{1/2} G_F^{-1/2}$.  The terms in the Lagrangian density,
involving $H$, finally are
\begin{eqnarray}
\frac{1}{2} (\partial_\mu H)^2 &-& \frac{1}{2} m^2_H H^2 - V_{SELF} (H) -
(\sqrt{2} G_F)^{1/2} \sum_f m_f \bar ff \nonumber\\[2mm]
&+& \frac{G_F}{\sqrt{2}} (2M^2_W
W^+_\mu W^{\mu^-} + M^2_Z Z^\mu Z_\mu) [2(G_F \sqrt{2})^{-1/2} H + H^2],
\nonumber
\end{eqnarray}
$V_{SELF} (H)$ containing the self-interaction terms of $H$.
Experimentally [8], from LEP 1, $m_H \ > \ 63.5$ GeV at present.

In the above we have discussed the electroweak MSM and its tree-level
relations in terms of three free parameters $g,g'$ and $v$.  Equivalently,
one may consider the measured constants $\alpha_{EM}, G_F$ and $M_Z$.
Among these the fine structure constant is known most precisely from
precision QED measurements:
$$
\alpha^{-1}_{EM} = 137.0359895 (61)
$$
Next, $G_F$ is known rather accurately from the charged current decay
process (Fig. 4) controlling the $\mu$-lifetime.  In fact we will call it
$G_\mu$ and use

\vspace {3.5 cm}
\begin{center}
Fig. 4. Muon decay diagrams
\end{center}
\begin{eqnarray}
\frac{1}{\tau_\mu} &=& \frac{G^2_\mu m^5_\mu}{192 \pi^3} (1 -
8m^2_e/m^2_\mu) \cdot \nonumber\\[2mm]
& & \cdot \left[1 + \frac{3}{5} m^2_\mu/M^2_W + (2\pi)^{-1}
\alpha_{EM} \left\{1 + 2(3\pi)^{-1} \alpha_{EM} \ell n(m_\mu/m_e)\right\}
(25/4 - \pi^2)\right], \nonumber
\end{eqnarray}
to deduce
$$
G_F = G_\mu = 1.166389(22) \times 10^{-5} ~{\rm (GeV)}^{-2}
$$
from the observed value of $\tau_\mu$.  There is also the crosssection
$\sigma$ for the neutral current induced elastic scattering process
$\nu_\mu e \rightarrow \nu_\mu e$, with incident neutrino energy $E_\nu$
in the lab frame, enabling us to define a low-energy neutral current weak
coupling:
$$
G_{NC} \equiv \left[\frac{2\pi\sigma}{m_e E_\nu} \left(1 - 4s^2_\theta +
\frac{16}{3} s^4_\theta\right)^{-1}\right]^{1/2} = \rho G_\mu.
\eqno (16)
$$
(16) can also be taken as a precise experimental definition of the
$\rho$-parameter which, as mentioned earlier, is unity at the tree level
for models of our interest. It may also be noted that (15) can be
recast as $\pi\alpha_{EM} = \sqrt{2} G_\mu M^2_W (1 - M^2_W/M^2_Z)$.  On
including radiative corrections, this generalizes to [9]
$$
\sqrt{2} G_\mu M^2_W (1 - M^2_W/M^2_Z) = \pi\alpha_{EM} (1 - \Delta
r)^{-1},
\eqno (17)
$$
where $\Delta r$ is a radiative parameter.

Finally, LEP 1, with an integrated luminosity of $\sim 150 ~{\rm
(pb)}^{-1}$, has given us a pretty accurate value of the $Z$-mass
currently at a $7 \times 10^{-5}$ precision level, namely
$$
M_Z = 91.187 (7) ~{\rm GeV}.
$$
Of course, there are additional ``dependent'' parameters in the MSM which
are directly known from experimental measurements, viz. the $W$-mass and
$s^2_\theta$.  The ``best'' values currently are
$$
\begin{array}{l}
M_W = 80.13 (28) ~{\rm GeV}, \\[2mm]
s^2_\theta = 0.2325 (5).
\end{array}
$$
Among the fermion masses $m_f$, from chiral symmetry considerations, it is
known that $5 ~{\rm MeV}<m_d<15 ~{\rm MeV}$ and $2 ~{\rm MeV}<m_u<8~{\rm
MeV}$ with $0.25<m_u/m_d<0.70$.  Moreover, similar considerations
extended to $K$ decays imply 100 MeV $\lsim M_S \lsim 300$ MeV.  Also,
charmonium
$J/\psi$ and bottomonium $\Upsilon$ considerations imply 1.3 GeV $\leq m_c
< 1.7$ GeV, $4.7 ~{\rm GeV} <  m_b  <  5.3$ GeV.  The current limit on the
top mass is 113 GeV $<m_t$.  For leptons $e$- and $\mu$-mass values are
as appear in Ref. [4] while the $\tau$-mass is now known to be $m_\tau =
1.777$ GeV.  The current upper limits on the neutrino masses are $m(\nu_e)
<7.2$ eV, $m(\nu_\mu) < 250$ keV and $m(\nu_\tau) < 31$ MeV.
\bigskip
\begin{itemize}
\item {\bf Basic LEP processes at the tree level} [10]
\end{itemize}
\bigskip

\nobreak
\noindent {\it Production and decay}
\bigskip

\nobreak
We start by giving the basic gauge boson-fermion pair vertices.  First
note that, for a vanishing fermion mass, left- and right-chirality for a
fermion corresponds to positive and negative helicity respectively.  The
second point to remember is that gauge vertices (unlike Yukawa ones) do
not change chirality since $\bar f\gamma_\mu f = \bar f_L \gamma_\mu f_L +
\bar f_R \gamma_\mu f_R$ and $\bar f \gamma_\mu \gamma_5 f = -f_L
\gamma_\mu f_L + \bar f_R \gamma_\mu f_R$.  We will find it convenient to
combine $v_f$ and $a_f$ of (13) into $\eta_{Lf}, \eta_{Rf}$:
$$
\begin{array}{l}
\eta_{Lf} = v_f + a_f,\\[2mm]
\eta_{Rf} = v_f - a_f.
\end{array}
$$
Ignoring the CKM matrix $V_{CKM}$ for quark flavor-mixing as a first
approximation, we can take the basic gauge-boson-fermion-pair vertices as
given in Fig. 5.
$$
\begin{array}{l}
{}~~~~~~~~~~~~~~~~~~~~~~~~~~~~~~g(2c_\theta)^{-1} \bar f\gamma_\mu (v_f -
a_f\gamma_5)f \\[2mm]
{}~~~~~~~~~~~~~~~~~~~~~~~~~~~~~~~~~~~~~~~~= \displaystyle {g \over
2c_\theta} (\eta_{Lf}
\bar f_L \gamma_\mu f_L + \eta_{Rf} \bar
f_R \gamma_\mu f_R) \\[8mm]
{}~~~~~~~~~~~~~~~~~~~~~~~~~~~~~~g(2\sqrt{2})^{-1} \bar f'\gamma_\mu (1 -
\gamma_5)f = g(\sqrt{2})^{-1}
\bar f'_L \gamma_\mu f_L \\[8mm]
{}~~~~~~~~~~~~~~~~~~~~~~~~~~~~~~eQ_f \bar f\gamma_\mu f = eQ_f (\bar f_L
\gamma_\mu f_L + \bar f_R \gamma_\mu f_R)
\end{array}
$$
\begin{center}
Fig. 5. Basic gauge-boson-fermion-pair vertices.
\end{center}

Suppose we generically describe both $W$- and $Z$-vertices by $\hat g \bar
f_2 \gamma_\mu (v - a\gamma_5)f_1$.  The tree level production cross
section for the process
$$
f_1 (p_1) \bar f_2(p_2) \rightarrow V(p), ~~~ V = W,Z
$$
with unpolarized beams and in the zero width approximation is given by
$$
\sigma(f_1\bar f_2 \rightarrow V) = \overline{\sum} |T_{1\bar 2}|^2 \pi
M^{-2}_V \delta(s - M^2_V),
\eqno (18)
$$
with $s = (p_1 + p_2)^2$ as the square of the CM energy.  As for the decay
$$
V \rightarrow f_1 \bar f_2,
$$
the partial width is
$$
\Gamma (V \rightarrow f_1 \bar f_2) = \frac{|\vec p|}{8\pi} M^{-2}_V
\overline{\sum} |T_{1\bar 2}|^2,
\eqno (19)
$$
where the CM fermion momentum $|\vec p| = \lambda(M^2_V,m^2_1,m^2_2) \equiv
(2m_V)^{-1} (m^4_V + m^4_1 + m^4_2 - 2m^2_V M^2_V - 2M^2_V m^2_2 - 2m^2_1
m^2_2)^{\frac{1}{2}} \simeq M_V/2$ for $M_V \gg m_{1,2}$.  In (18) and
(19) $T_{1\bar 2}$ is the transition matrix element $\hat g\bar v(p_2)
\gamma^\mu (v - a\gamma_5) u(p_1) \epsilon^\star_\mu$ in standard notation
(our normalization is $\bar uu = \bar vv = 2m$) and $\sum$ stands for the
summation over all colors and spins while the bar on top implies division
by initial spin and color factors.  Thus
\begin{eqnarray}
\overline{\sum} |T_{1\bar 2}|^2 = \hat g^2 M^2_V N^c_f \Big[(v^2 +
a^2) \Big\{1 - \frac{1}{2} m^{-2}_V (m^2_1 &+& m^2_2) - \frac{1}{2}
m^{-4}_V (m^2_1 - m^2_2)^2\Big\} \nonumber \\[2mm]
&+& 3(v^2 - a^2)M^{-2}_V m_1 m_2\Big]. \nonumber
\end{eqnarray}
The color factor $N^c_f$ is $3$ for quarks and $1$ for leptons.

For light fermions, substituting the right value of $\hat g$ for $CC$ and
$NC$ interactions, we have
$$
\Gamma (W \rightarrow f_1\bar f_2) = \displaystyle {\sqrt{2} \over 12\pi}
G_\mu M^3_W N^C_f |V^{12}|^2,
\eqno (20a)
$$
$$
\Gamma (Z \rightarrow f\bar f) = \displaystyle {\sqrt{2} \over 12\pi}
G_\mu M^3_Z N^C_f
(v^2_f + a^2_f) = \frac{\alpha_{EM}}{6s^2_\theta c^2_\theta} (T_{3f} -
s^2_\theta Q_f)^2 N^C_f,
\eqno (20b)
$$
where $V^{12}$ is unity when $f$ is a lepton and equals $V^{12}_{CKM}$
when $f$ is a quark.  Interestingly, in the limit $s_\theta \rightarrow 0$,
$M_Z \rightarrow M_W$ and one has the sumrule $\Gamma (W \rightarrow f_1\bar
f_2) \rightarrow \Gamma (Z \rightarrow f_1 \bar f_1) + \Gamma (Z \rightarrow
f_2 \bar f_2)$.  Three remarks are in order.  (1) Since the top is heavier
than 113 GeV, the decays $W
\rightarrow tb$, $Z \rightarrow t\bar t$ are not possible, (2) LEP 1 has
excluded any extra heavy fermion upto a mass of $\sim 45$ GeV and (3) the
heaviest known fermion, viz. the $b$-quark, weighs only 4.8 GeV, so that
fermion mass terms in $W$- and $Z$-decay are mostly neglected.  We can consider
$Z$ decays into particle-antiparticle pairs not only of charged leptons
but also of neutrinos $\nu_e,\nu_\mu$ or $\nu_\tau$ and define $\Gamma (Z
\rightarrow$ invisible) $= 3\Gamma (Z \rightarrow \nu\bar \nu)$; similarly
we can write $\Gamma (Z \rightarrow$ hadrons) $= \displaystyle \sum_{5q}
\Gamma (Z \rightarrow q\bar q)$ where $\displaystyle \sum_{5q}$ is over
$u,d,s,c$ and $b$.  Finally,
$$
\Gamma^{tot}_Z \simeq \sum_f \Gamma(Z \rightarrow f\bar f) = \sum_f
\Gamma(Z \rightarrow \nu\bar\nu) (1 - 4|Q_f|s^2_\theta + 8Q^4_f
s^4_\theta).
\eqno (21)
$$
It may be noted that, $m_H$ being $> \ 63.5$ GeV, any contribution to
$\Gamma^{tot}_Z$ from a final state containing $H$ would be quite small.
One can define $\Gamma (Z \rightarrow$ invisible) $= N_\nu \Gamma(Z
\rightarrow \nu\bar \nu)$ and experimentally determine $N_\nu$ from LEP 1.
The current result is $N_\nu = 3 \pm 0.04$.

Let us move on to $Z$-production in $e^+e^-$ collisions.  We stick with
the light fermion approximation.  Thus, from (19), with the notation
$\Gamma^f_Z$ for $\Gamma (Z \rightarrow f\bar f)$,
$$
\sum |T_{1\bar 2}|^2 = 48 \pi M^2_V M^{-1}_Z \Gamma^f_Z.
$$
(18) now implies that -- in the zero-width approximation --
$$
\sigma(e^+e^- \rightarrow Z) = 12\pi^2 M^{-1}_Z \Gamma^e_Z \delta (s -
M^2_Z),
$$
a formula -- which for finite widths -- has to be modified via
$$
\pi \delta (s - M^2_Z) \rightarrow M_Z \Gamma^{tot}_Z [(s - M^2_Z)^2 +
M^2_Z (\Gamma^{tot}_Z)^2]^{-1}
$$
to the Breit-Wigner form
$$
\sigma (e^+e^- \rightarrow Z) = 12\pi \Gamma^e_Z \Gamma^{tot}_Z [(s -
M^2_Z)^2 + M^2_Z (\Gamma^{tot}_Z)^2]^{-1}.
\eqno (22)
$$
Near the $Z$-resonance,
\begin{eqnarray}
\sigma(e^+e^- \rightarrow Z \rightarrow f\bar f) &=& \sigma(e^+e^-
\rightarrow Z) \Gamma^f_Z (\Gamma^{tot}_Z)^{-1} \nonumber\\[2mm]
&=& 12\pi \Gamma^e_Z
\Gamma^f_Z \left[(s - M^2_Z)^2 + M^2_Z (\Gamma^{tot}_Z)^2\right]^{-1}.
{}~~~~~~~~~~(23) \nonumber
\end{eqnarray}

\noindent {\it Scattering cross sections}
\bigskip

\nobreak
Consider the process
$$
e^- (p_1) e^+ (p_2) \longrightarrow f(q_1) \bar f(q_2)
$$
in the Born approximation involving $\gamma^\star$ and $Z^\star$ exchange
in the $s$-channel, as shown in Fig. 6.
Because of chirality-conservation at each vertex, there are four
possible independent transition amplitudes: $T_{h_e} T_{h_f} =
T_{LL},T_{LR},T_{RL}$, $T_{RR}$.  Here $h_e,h_f$ refer the handedness of the
$e,f$ being $L$ or $R$.  Thus we have

\newpage

\vspace* {3.5 cm}

\begin{center}
Fig. 6. Tree diagrams for $e^+e^- \rightarrow f\bar f$.
\end{center}

$$
\frac{d\sigma}{d\cos\Theta} = \frac{s}{48\pi} N^C_f \sum_{spins}
|T_{h_eh_f}|^2,
\eqno (24)
$$
with $\Theta$ as the $CM$ scattering angle.  Employing the $\eta$-notation
introduced earlier,
\begin{eqnarray}
|T_{h_eh_f}|^2 = \frac{3}{8} (1 \pm \cos\Theta)^2 \Big|\eta_{h_ee}
\eta_{h_ff} \sqrt{2} G_\mu M^2_Z (s - M^2_Z &+& iM_Z\Gamma_Z)^{-1}
\nonumber\\[2mm]
&+& 4\pi \alpha Q_e Q_f s^{-1}\Big|^2, \nonumber
\end{eqnarray}
with the $+(-)$ sign being for $T_{LL}$ and $T_{RR}$ $(T_{LR}$ and
$T_{RL})$.

(24) can now be rewritten as
$$
\frac{d\sigma}{d\cos\Theta} = \frac{d\sigma^\gamma}{d\cos\Theta} +
\frac{d\sigma^Z}{d\cos\Theta} + \frac{d\sigma^{\gamma Z}}{d\cos\Theta},
$$
the individual pieces standing for the pure $QED$, the pure weak and the
electroweak interference terms respectively.  The first and the second
pieces dominate at low energies and near the $Z$-resonance respectively.
To show these pieces in detail, we define
$$
{\cal R} (s) \equiv s[s - M^2_Z + iM_Z \Gamma^{tot}_Z]^{-1}
$$
and write
$$
\begin{array}{l}
{\displaystyle\frac{d\sigma^\gamma}{d\cos\Theta}} = {\displaystyle
\frac{\pi\alpha^2_{EM}}{2s}} Q^2_f
N^C_f (1 + \cos^2\Theta),\\[2mm]
{\displaystyle \frac{d\sigma^Z}{d\cos\Theta}} = {\displaystyle
\frac{G^2_\mu M^4_Z}{16\pi s}} N^C_f |{\cal
R}(s)|^2 [(v^2_e + a^2_e) (v^2_f + a^2_f) (1+\cos^2\Theta) +
8a_ea_fv_ev_f\cos\Theta], \\[2mm]
{\displaystyle \frac{d\sigma^{\gamma Z}}{d\cos\Theta}} = - {\displaystyle
\frac{\alpha_{EM} \sqrt{2}
G_\mu M^2_Z}{4s}} Q_f N^C_f ~{\rm Re}~{\cal R}(s) [v_e v_f (1 +
\cos^2\Theta) + 2a_ea_f \cos\Theta]. \\[1mm]
\end{array}
\eqno (25)
$$
It is noteworthy that the terms linear in $\cos\Theta$ in the RHS of (25)
would vanish if there was no axial vector coupling in weak interactions.

The total cross section splits into pure vector and axial vector pieces:
$$
\begin{array}{l}
\sigma_{f\bar f} = \int^1_{-1} d\cos\Theta {\displaystyle
\frac{d\sigma}{d\cos\Theta}} =
\sigma^{VV}_{f\bar f} + \sigma^{AA}_{f\bar f},\\[2mm]
\sigma^{VV}_{f\bar f} = {\displaystyle \frac{4\pi\alpha^2_{EM}}{3s}} Q^2_f
N^C_f -
{\displaystyle \frac{2\alpha_{EM} \sqrt{2} G_\mu M^2_Z}{3s}} Q_f N^C_f
{}~{\rm Re}~ {\cal R} (s) v_e v_f \\[2mm]
{}~~~~~~~~~~~~~~~~~~~~~~~~~~~~~ + {\displaystyle \frac{G^2_\mu M^4_Z}{6\pi s}}
N^C_f |{\cal R}|^2 (v^2_e + a^2_e)v^2_f, \\[2mm]
\sigma^{AA}_{f\bar f} = {\displaystyle \frac{G^2_\mu M^4_Z}{6\pi s}} N^C_f
|{\cal R}|^2 (v^2_e + a^2_e) a^2_f.
\end{array}
\eqno (26)
$$
Near the $Z$-resonance the total cross section can be written as
$$
\sigma_{f\bar f} \simeq \sigma^Z_{f\bar f} \left[1 + \frac{8\pi\alpha_{EM}
Q_e Q_f}{\sqrt{2} G_\mu M^2_Z} \frac{v_e v_f} {(v^2_e + a^2_e) (v^2_f +
a^2_f)} \frac{s - M^2_Z}{s}\right] + \sigma^\gamma_{f\bar f}.
\eqno (27)
$$
On resonance, where $s = M^2_Z$, the background term
$$
\sigma^\gamma_{f\bar f} = \frac{4\pi\alpha^2_{EM} Q^2_f}{3M^2_Z} N^C_f
$$
is a $< \ 1\%$ correction to the dominant term
$$
\sigma^Z_{f\bar f} = \frac{3G^2_\mu M^4_Z}{4\pi [\Gamma^{tot}_Z]^2} N^C_f
(v^2_e + a^2_e) (v^2_f + a^2_f).
\eqno (28)
$$
It should be pointed out that the detailed fitting of near-resonance $Z$
lineshape requires higher order corrections.  In particular, these oblige
one to use an energy-dependent width $\Gamma^{tot}_Z (s)$.
\bigskip

\noindent {\it Asymmetries}

\bigskip
\nobreak
The differential cross section, discussed above, has a term that is even
in $\cos\Theta$ and one that is odd in $\cos\Theta$
\begin{eqnarray}
\frac{d\sigma_{f\bar f}}{d\cos\Theta} &=& \sigma_{f\bar f} (s) \frac{3}{8}
(1 + \cos^2\Theta) + \cos\Theta  \Big[\frac{G^2_\mu
M^4_Z}{2\pi s} N^C_f |{\cal R}|^2 a_e a_f v_e v_f\nonumber \\[2mm]
& & - \frac{\alpha_{EM} \sqrt{2} G_\mu
M^2_Z} {2s} Q_f N^C_f ({\rm Re}~{\cal R}) a_e a_f\Big]
{}~~~~~~~~~~~~~~~~~~~~~~~~~~~~~ (29)
\nonumber\\[2mm]
& \equiv & \frac{3}{8} \sigma_{f\bar f} (s) (1 + \cos^2\Theta) +
\Delta^0_{f\bar f} (s) \cos\Theta. \nonumber
\end{eqnarray}
Because of the odd term, a forward-backward asymmetry, viz.
$$
{\cal A}^{FB}_{f\bar f} (s,\cos\Theta) = \frac{d\sigma_{f\bar f} (\Theta) -
d\sigma_{f\bar f} (\pi - \Theta)} {d\sigma_{f\bar f} (\Theta) +
d\sigma_{f\bar f} (\pi - \Theta)} = \frac{8}{3} \frac{\Delta_{f\bar f}
(s)} {\sigma_{f\bar f} (s)} \frac{\cos\Theta}{1 + \cos^2\Theta}
\eqno (30)
$$
is generated.  There is an angular-integrated version of the
forward-backward asymmetry, namely
$$
A^{FB}_{f\bar f} (s) = \frac{(\int^1_0 - \int^0_{-1}) d\cos\Theta
{\displaystyle \frac{d\sigma_{f\bar f}}{d\cos\Theta}}} {\int^1_{-1}
d\cos\Theta {\displaystyle \frac{d\sigma_{f\bar
f}}{d\cos\Theta}}} = \frac{\Delta_{f\bar f} (s)} {\sigma_{f\bar f} (s)}.
\eqno (31)
$$
This is, of course, easier to measure because of the higher statistics.
The general tree-level expressions for the forward-backward asymmetry
appear in eqs. (29)-(31), but it is useful to consider two special energy
domains:

(1) Small $s \ll M^2_Z$.

\noindent Now
$$
\sigma_{f\bar f} \simeq \frac{4\pi\alpha^2_{EM}} {3s} \left[Q^2_f N^C_f +
\frac{\sqrt{2} G_\mu}{2\pi \alpha_{EM}} N^C_f Q_f \frac{v_e v_f} {1 -
sM^{-2}_Z}\right].
\eqno (32)
$$
In (32) the factor outside is the $QED$ ``point cross section'' used to
normalize $\sigma ~(e^+e^- \rightarrow$ hadrons).  Furthermore,
$$
A^{FB}_{f\bar f} (s) \simeq \frac{3}{8} \frac{a_e a_f} {Q_f} \frac{\sqrt{2}
G_\mu} {\pi \alpha_{EM}} \frac{s} {1 - sM^{-2}_Z}
\eqno (33)
$$
which vanishes as $s \rightarrow 0$.

(2) For $s \simeq M^2_Z$ we find
$$
A_{f\bar f}^{FB} (M^2_Z) = \frac {\Delta^0_{f\bar f} (M^2_Z)} {\sigma_{f\bar f}
(M^2_Z)} = \frac {3} {4} \frac{2v_e a_e} {v^2_e + a^2_e} \frac {2v_f a_f}
{v^2_f + a^2_f} = \frac {3} {4} \frac{\eta^2_{Le} - \eta^2_{Re}}
{\eta^2_{Le} + \eta^2_{Re}} \frac{\eta^2_{Lf} - \eta^2_{Rf}}
{\eta^2_{Lf} + \eta^2_{Rf}}.
\eqno (34)
$$
It is important to note that
$$
\frac {\eta^2_{L\ell} - \eta^2_{R\ell}} {\eta^2_{L\ell} + \eta^2_{R\ell}} =
\frac {2\xi} {1 + \xi^2},
$$
with $\xi = 1 - 4s^2_\theta \simeq 0.10$ for $\ell = e,\mu,\tau$.  Thus
$$
A^{FB}_{\mu\bar \mu} (M^2_Z) = \frac {3\xi^2} {(1 + \xi^2)^2}.
\eqno (35)
$$
Near $s = M^2_Z$,
$$
A^{FB}_{f\bar f} (s) \simeq \frac {3v_e a_e v_f a_f} {(v^2_e + a^2_e)
(v^2_f + a^2_f)} - \frac {3} {2\pi} \frac {\sqrt{2} G_\mu M^2_Z}
{\alpha_{EM}} (1 - M^2_Z s^{-1}) \frac {a_e a_f} {(v^2_e + a^2_e) (v^2_f +
a^2_f)}.\\[2mm]
\eqno (35a)
$$

Another interesting asymmetry concerns the polarization of the final state
fermion $f$ (such as a $\tau$):
$$
A^f_{pol} \equiv \frac {\sigma (e^+e^- \rightarrow f_L \bar f) - \sigma
(e^+e^- \rightarrow f_R \bar f)} {\sigma (e^+e^- \rightarrow f_L \bar f) +
\sigma (e^+ e^- \rightarrow f_R \bar f)}.
$$
In fact, on the $Z$, this yields
$$
A^f_{pol} (M^2_Z) = \frac {\eta^2_{Lf} - \eta^2_{Rf}} {\eta^2_{Lf} +
\eta^2_{Rf}},
$$
independently of initial couplings.  It may be noted that $A^f_{FB}
(M^2_Z) = \frac {3}
{4} A^e_{LR} (M^2_Z) A^f_{LR} (M^2_Z)$.  The experimental measurement of
$A^f_{pol}$ is most feasible for $f = \tau$ since the $\tau$-polarization
can be measured from the decays $\tau \rightarrow \pi\nu,\rho\nu,a_1\nu$
with the subsequent decays $\rho \rightarrow \pi\pi$, $a_1 \rightarrow 3\pi$ as
well as from $\tau \rightarrow \nu +$ jets.  For $e^+e^- \rightarrow
\tau^+\tau^-$,
$$
A^\tau_{pol} (M^2_Z) = \frac {2\xi} {1 + \xi^2}.
$$

There are additional asymmetries involving polarized beams which are of
great theoretical interest and await futuristic experiments.  But since
clean polarized high energy $e^\pm$ beams will not be available for some
time, we do not discuss them in detail except to define
$$
A_{LR} \equiv \frac {\sigma (e^-_L e^+ \rightarrow f\bar f) - \sigma
(e^-_R e^+ \rightarrow f\bar f)} {\sigma (e^-_L e^+ \rightarrow f\bar f) +
\sigma (e_R e^+ \rightarrow f\bar f)}
$$
and quote
$$
A_{LR} (M^2_Z) = \frac {\eta^2_{Le} - \eta^2_{Re}} {\eta^2_{Le} +
\eta^2_{Re}} = \frac {2\xi} {1 + \xi^2}.
$$

\newpage

\begin{itemize}

\item {\bf 1-loop radiative corrections in the on-shell renormalization
scheme}

\end{itemize}

What are our independent couplings?  First, the fine structure constant
$\alpha_{EM}$ and the mass $M_Z$.  These define an on-shell QED-like [11]
renormalization scheme [12].  Then there is the $\mu$-decay coupling
constant given at the tree level by
$$
\sqrt{2} G_\mu = {1 \over v^2} = {\pi\alpha_{EM} \over M^2_W s^2_\theta},
{}~~~s^2_\theta = 1 - M^2_W M^{-2}_Z.
$$
Note that the gauge couplings $g = \sqrt {4\pi\alpha_{EM}}/s_\theta$, $g'
= \sqrt {4\pi\alpha_{EM}}/c_\theta$ are in the category of dependent
parameters.  Next, one has to renormalize these parameters.  Finally come
the field or wave function renormalizations.

The input parameters in the true bare Lagrangian are $M^2_{Wb}$,
$M^2_{Zb}$, $\alpha_b$ and $G_{\mu b}$ where the subscript $b$ signifies
bare values.  After the 1-loop correction, they
become cutoff dependent (i.e. infinite).  Then they have to be
reparametrized in terms of the corresponding finite physical parameters by
additive infinite renormalization constants.  Thus,
$$
\begin{array}{l}
M^2_{W,Zb} = M^2_{W,Z} \left(1 + \displaystyle {\delta M^2_{W,Z} \over
M^2_{W,Z}} \right), \\[2mm]
\alpha_{EMb} = \alpha_{EM} \left(1 + \displaystyle {\delta \alpha_{EM} \over
\alpha_{EM}}\right), \\[2mm]
s^2_{\theta b} = s^2_\theta \left(1 + \displaystyle {\delta s^2_\theta \over
s^2_\theta}\right), \\[2mm]
G_{\mu b} = G_\mu \left(1 + \displaystyle {\delta G_\mu \over G_\mu}\right).
\end{array}
\eqno (36)
$$
Finally, we have
$$
{\delta G_\mu \over G_\mu} = {\delta \alpha_{EM} \over \alpha_{EM}} -
{\delta M^2_W \over M^2_W} - {\delta s^2_\theta \over s^2_\theta}.
\eqno (37)
$$
Turning to field renormalization, we ignore the infrared problem due to
soft photons in QED.  That is an old subject and is understood in terms of
standard
textbook techniques.  We simply attribute a mass $m_\gamma$ to the photon
as an infrared regulator and require that the limit $m_\gamma \rightarrow
0$ must exist for all observable quantities.

For fields, then, write the renormalized objects as
$$
\begin{array}{l}
V_{\mu b} = \sqrt {Z_V} V_{\mu r} ~~~ (V = A,W^\pm,Z), \\[2mm]
f_b = \sqrt {Z_f} f_r, \\[2mm]
H_b = \sqrt {Z_H} H_r,
\end{array}
\eqno (38)
$$
where the wave function renormalization constants $Z$ are fixed by the
condition that {\it propagators of the renormalized fields have unit
residues at their poles}.  To leading order, $Z_i = 1$ and we may write
$$
Z_i = 1 + \delta Z_i; ~~~\sqrt{Z_i} \simeq 1 + {1 \over 2} \delta Z_i +
\cdots .
$$

The actual renormalization procedure of a physical amplitude can be done
as follows.  First, perform the parameter shifts and field
renormalizations to 1-loop by expanding upto linear order.  Thus, for the
$f\bar f V$ vertices substitute
$$
\begin{array}{l}
{}~~~~~ (bare) ~~~~~~~~~~~~~~~~~~~~~~~~~~~~~~~~~~~~~~~~~~~~~~
(renormalized) \\[2mm]
eQ_f \gamma_\mu  ~~~~~~~~~~~~~~~~~~~~~~~~~~~~~~~~~~~ \rightarrow eQ_f
\gamma_\mu \left(1 + {1
\over 2} \delta Z_A + \delta Z_f + \displaystyle {\delta e \over e}\right)
\\[4mm]
\sqrt {2G_\mu} M_Z \gamma^\mu [T_{3f} (1 - \gamma_5) - 2s^2_\theta Q_f]
\rightarrow \sqrt {2G_\mu} M_Z \gamma^\mu [T_3(1 - \gamma_5) \\[2mm]
{}~~~~~~~~~~~~~~~~~~~~~~~~~~~~~~~~~~~~~~~~~~~~~~~~- 2Q_f
s^2_\theta (1 + \delta s^2_\theta/s^2_\theta)] \cdot \\[2.5mm]
{}~~~~~~~~~~~~~~~~~~~~~~~~~~~~~~~~~~~~~~~~~~~~~~~~\left(1 + {1\over2} \delta
Z_Z + \delta Z_f + \displaystyle {1\over2} \displaystyle {\delta M^2_Z
\over M^2_Z} + {1\over2} \displaystyle {\delta G_\mu \over G_\mu}\right)
\\[4.5mm] \sqrt {2\sqrt{2} G_\mu} M_W \gamma^\mu (1 - \gamma_5)
{}~~~~~~~~~~~~~\rightarrow \sqrt
{2\sqrt{2} G_\mu} M_W \gamma^\mu (1 - \gamma_5) \Bigg(1 + {1\over2} \delta
Z_W\\[4mm]
{}~~~~~~~~~~~~~~~~~~~~~~~~~~~~~~~~~~~~~~~~~~~~~~~~~ + {1\over2} \delta Z_{f_1}
+ {1\over2} \delta Z_{f_2} + \displaystyle {1\over2}
\displaystyle {\delta M^2_W \over M^2_W} + \displaystyle {1\over2}
\displaystyle {\delta G_\mu \over G_\mu}\Bigg).
\end{array}
$$
Analogous substitutions have to be made for other vertices.

Next, the mass counter terms and the wave function factors have to be
introduced.  These are determined by the physical tranverse parts of the
vector boson self-energy:
$$
D^{\mu\nu}_{VV} (q^2) = {-i\eta^{\mu\nu} \over q^2 - M^2_V - \Pi_{VV}
(q^2)} + q^\mu q^\nu ~{\rm term}.
\eqno (39)
$$
In (39) the $\eta^{\mu\nu}$-term signifies the tranverse part whereas the
unspecified $q^\mu q^\nu$ term stands for the longitudinal part.
{}~~$D^{\mu\nu}_{VV}$ ~arises ~from ~a ~sum ~of

\vspace*{4 cm}

\begin{center}
Fig. 7. Perturbation series for vector boson self-energies.
\end{center}

\noindent  the bare propagator and vacuum polarization bubbles as
shown in Fig. 7.  Here $\Pi_{VV}$ is defined as the coefficient of
$(-i\eta^{\mu\nu})$ in the 1-loop vector boson self-energies:
$$
\Pi^{\mu\nu} (q) = i\Pi_{VV} (q^2) \eta^{\mu\nu} + q^\mu q^\nu ~{\rm term}.
$$
Thus
\begin{eqnarray}
{-i\eta^{\mu\nu} \over q^2 - M^2_V} &=& {-i\eta^{\mu\nu} \over q^2 - M^2_V
- \Pi_{VV} (q^2)} \cdot \nonumber \\[2mm]
& & \left[1 + \Pi_{VV} (q^2) {1 \over q^2 - M^2_V} +
\Pi_{VV} (q^2) {1 \over q^2 - M^2_V} \Pi_{VV} {1 \over q^2 - M^2_V} +
\cdots \right]. \nonumber
\end{eqnarray}

Since the vector boson self-energy is quadratically divergent, two
subtractions (chosen on-shell) are needed so that
$$
\Pi_{VVr} (q^2) = \Pi_{VV} (q^2) - \Pi_{VV} (M^2_V) - (q^2 - M^2_V)
{d\Pi_{VV} (M^2_V) \over dq^2} + {\rm higher~orders}.
$$
(Note that the unrenormalized $\Pi$-function is a divergent,
regulator-dependent quantity).  The tranverse part of the free inverse
propagator is changed as under
\begin{eqnarray}
i\eta^{\mu\nu} (q^2 - M^2_{Vb}) &\rightarrow& i\eta^{\mu\nu} Z_V (q^2 -
M^2_V - \delta M^2_V) \nonumber \\[2mm]
&=& i\eta^{\mu\nu} [q^2 - M^2_V - \delta M^2_V + \delta Z_V (q^2 - M^2_V)
+ \cdots ], \nonumber
\end{eqnarray}
where $\delta Z_V = Z_V - 1$.  Diagrammatically, this entry of the counter
terms is shown in Fig. 8.

\vspace* {4 cm}
\begin{center}
Fig. 8. Composition of renormalized vector boson propagator.
\end{center}

On mass-shell renormalization now implies that the transverse
part of the vector boson renormalized self-energy and its derivative vanish
at $q^2 = M^2_V$.  This means
$$
\begin{array}{l}
\delta M^2_V = - {\rm Re}~ \Pi_{VV} (M^2_V) \\[2mm]
\delta Z_V \equiv Z_V - 1 = + {\rm Re}~ \displaystyle {d\Pi_{VV} \over dq}
(M_V^2).
\end{array}
\eqno (40)
$$
Since $Z,W$ are unstable, the self-energy has an imaginary part too:\\
${\rm Im}~\Pi_{VV}$ $(M^2_V) \equiv M_V \Gamma_V$ giving the total width
of $V$.
We will henceforth drop the prefix ${\rm Re}$.

The above treatment, though exactly valid for $W$, has to be modified both
for the photon and the $Z$ because of $\gamma - Z$ mixing.  In place of
(39), one would now have a $2 \times 2$ $\gamma - Z$ symmetric inverse
propagator matrix [13]
$$
\hat D^{-1} = \left(\matrix{q^2 - \Pi_{\gamma\gamma} (q^2) & -\Pi_{\gamma
Z} (q^2) \cr -\Pi_{\gamma Z} (q^2) & q^2 - M^2_Z - \Pi_{ZZ} (q^2)}\right)
\eqno (41)
$$
Taking inverse and keeping only linear terms in $\Pi$ (consistent to
1-loop), we have
$$
\begin{array}{l}
D_{\gamma\gamma} \simeq \displaystyle {1 \over q^2} \left[1 +
\displaystyle {\Pi_{\gamma\gamma} (q^2)
\over q^2}\right], \\[2mm]
D_{\gamma Z} \simeq \displaystyle {\Pi_{\gamma Z} (q^2) \over q^2 (q^2 -
M^2_Z)},
\\[2mm] D_{ZZ} \simeq \displaystyle {1 \over q^2 - M^2_Z} \left[1 +
{\Pi_{ZZ} (q^2) \over q^2 - M^2_Z}\right].
\end{array}
\eqno (42)
$$
(We shall have more to say about these pi-functions later on.)  All the
vector boson self-energy diagrams in a general gauge are given in Fig. 9.

\vspace* {3.5 cm}
\begin{center}
Fig. 9. Vector boson self-energies in a general gauge.
\end{center}

\noindent Of course, the tadpole contributions drop out of renormalized
quantities.

Finally, we have to come to the fermions.  The fermion propagator
renormalization is straightforward (Fig. 10) except that it has to be
separate

\vspace* {3 cm}

\begin{center}
Fig. 10. Fermion self-energy diagrams.
\end{center}

\noindent for the different chiral components.  Equivalently, we can write
$$
\delta Z_f = \delta Z_{vf} + \delta Z_{af} \gamma_5.
\eqno (43)
$$
Coming to the fermionic electric charge (Fig. 11), the condition is that,
to

\vspace* {3 cm}
\begin{center}
Fig. 11. Fermion electric charge renormalization.
\end{center}

\noindent 1-loop, the $EM$ coefficient of the $\gamma_\mu$ vertex must
lead in the zero photon energy limit $(q^2 = 0)$ to the renormalized
charge $e$.  The vertex correction diagrams to 1-loop are shown in Fig. 12.

\vspace* {3.5 cm}
\begin{center}
Fig. 12. 1-loop vertex corrections to fermion electric charge.
\end{center}

\noindent The renormalization condition, together with the $EM$
Ward-identity following from current-conservation $\partial_\mu J^{\mu EM}
= 0$, leads to a constraint linking some vertex correction contributions
to self-energy ones, but we shall not go into these details.  They may be
found in the review article by Jegerlehner [12].

We want to end this discussion with remarks on the radiative corrections
to $G_\mu$, measured in the decay $\mu \rightarrow e\bar \nu_e \nu_\mu$.
The effective pointlike interaction is
$$
-4 \displaystyle \frac{G_\mu}{\sqrt{2}} \bar u_{\nu_\mu} \gamma_\alpha
\frac{1}{2} (1 -\gamma_5) u_\mu \bar u_e \gamma^\alpha \frac{1}{2} (1 -
\gamma_5) v_{\nu_e}.
$$
The 1-loop radiative corrections come from vector boson self-energy
diagrams, vertex corrections and box diagrams.  These are schematically
enumerated along with the tree graph in Fig. 13. In more specific detail,

\vspace* {4 cm}
\begin{center}
Fig. 13. Schematic enumeration $\mu$-decay diagrams to 1-loop.
\end{center}

\noindent  the vertex corrections (top line) and the box
contributions are shown in Fig. 14.

\vspace* {5 cm}
\begin{center}
Fig. 14. Vertex corrections and box contributions to $\mu$-decay at
1-loop.
\end{center}

To one loop, one can approximate $[M^2_W + \Pi_{WWr} (0)]^{-1}$
to $M^{-2}_W [1 - M^{-2}_W \Pi_{WWr} (0)$.  Thus Fig. 13, rewritten as an
equation, reads
$$
\displaystyle \frac{G_\mu} {\sqrt{2}} = \displaystyle \frac{e^2_b}
{8s^2_{\theta b} M^2_{Wb}} \left\{1 - \displaystyle \frac{\Pi_{WWr} (0)}
{M^2_W} + \delta_{VERTEX} + \delta_{BOX}\right\}.
\eqno (44)
$$
Detailed expressions for $\delta_{VERTEX}$ and $\delta_{BOX}$ may be found
in Refs. [12] and [14].  Rewriting all bare quantities in terms of the
corresponding
renormalized quantities and employing the trick $\delta
c_\theta/s^2_\theta = c^2_\theta s^{-2}_\theta \delta
c^2_\theta/c^2_\theta$, we can write
\begin{eqnarray}
\displaystyle \frac{G_\mu} {\sqrt{2}} &=& \displaystyle\frac{e^2}
{8s^2_\theta M^2_W} \Bigg\{1 + \displaystyle \frac{\delta\alpha_{EM}}
{\alpha_{EM}} - \displaystyle \frac{c^2_\theta} {s^2_\theta}
\left(\displaystyle \frac{\delta M^2_Z} {M^2_Z} - \displaystyle\frac
{\delta M^2_W} {M^2_W}\right) - \displaystyle\frac {\delta M^2_W} {M^2_W}
\nonumber \\[2mm]
& & ~~~~~~~~~~~~~~~ - \displaystyle \frac {\Pi_{WWr} (0)} {M^2_W} +
\delta_{VERTEX} + \delta_{BOX}\Bigg\}, \nonumber \\[2mm]
&=& \displaystyle \frac{\pi\alpha_{EM}} {2s^2_\theta M^2_W} (1 + \Delta
r), ~~~~~~~~~~~~~~~~~~~~~~~~~~~~~~~~~~~~~~~~~~~~~~~~~~~~~~~~ (45) \nonumber
\end{eqnarray}
where $\Delta r = (\Delta r)_{SE} + (\Delta r)_{VERTEX} + (\Delta
r)_{BOX}$.

We will not go into the detailed technical calculation of $\Delta r$ for
which the best reference is Hollick's article [12].  Suffice it to say
that the vertex corections and box graph terms are an order of
magnitude smaller and that the main contribution to $\Delta r$ comes from
the various self-energy sources.  This is a generic property of all $EW$
radiative corrections of interest except the $Zb\bar b$ vertex.  These
vector boson self-energy corrections, which form a gauge-invariant subset,
have been called [16] {\it oblique.}  As an example let us look (Fig. 15)
at the vacuum polarization contribution to the photon self-

\vspace* {4 cm}
\begin{center}
Fig. 15. Fermion-summed photon vacuum polarization.
\end{center}

\noindent  energy, summed over all fermions.  This leads to
$$
\displaystyle\frac{\Delta \alpha_{EM}} {\alpha_{EM}} = \displaystyle \frac
{1} {3\pi} \sum_f Q^2_f N^C_f \left(\ell n \displaystyle \frac {M^2_Z}
{m^2_f} - \frac{5}{3}\right),
\eqno (46)
$$
where $\displaystyle \sum_f$ covers both quarks and leptons.  The RHS of
(46) has a leptonic contribution which is straightforward.  The nontop
hadronic contribution should not be perturbatively calculated because of
nonperturbative QCD effects and is best given by the dispersion integral
$$
- \displaystyle \frac {M^2_Z} {3\pi} ~{\rm Re}~ \int^\infty_{4m^2_\pi} ds
(s - M^2_Z + i\epsilon)^{-1} [\sigma_{e^+e^- \rightarrow \mu^+\mu^-}
(s)]^{-1} [\sigma_{e^+e^- \rightarrow ~{\rm hadrons}} (s)].
$$
On the other hand, the top contribution, owing to the high value of
$m_{top}$, is perturbative.  However, there is another source of the top
contribution, namely the $W$ self-energy (Fig. 16).

\vspace* {2.5 cm}
\begin{center}
Fig. 16. Top contribution to $W$ self-energy.
\end{center}

\noindent All told, the top-contribution to $\Delta r$ is:
$$
\Delta r^{top} = - \displaystyle \frac {\sqrt{2} G_\mu M^2_W} {16\pi^2}
\left\{3 \displaystyle \frac {c^2_\theta} {s^2_\theta} \displaystyle \frac
{m^2_t} {m^2_W} + 2\left(\displaystyle \frac {c^2_\theta} {s^2_\theta} -
\frac {1} {3}\right)\ell n \displaystyle \frac {m^2_t} {M^2_W} + \frac {4}
{3} \ell n c^2_\theta + \displaystyle \frac {c^2_\theta} {s^2_\theta} -
\frac {7} {9}\right\}
\eqno (47)
$$
and has a piece that is quadratic in $m_t$.  The Higgs contribution,
coming basically from the diagrams of Fig. 17, is
$$
\Delta r^{Higgs} = \displaystyle \frac {\sqrt {2} G_\mu M^2_W} {16\pi^2}
\cdot \displaystyle {11} {3} \left(\ell n \displaystyle \frac {m^2_H}
{M^2_W} - \frac {5} {6}\right).
$$

\vspace* {4 cm}
\begin{center}
Fig. 17. Higgs contributions to vector boson self-energies.
\end{center}
\bigskip

\begin{itemize}
\item {\bf STAR SCHEME AND OBLIQUE CORRECTIONS}
\end{itemize}

\bigskip
\nobreak
\noindent {\it Vacuum Polarizations}

\bigskip
\nobreak
Though the on-mass-shell renormalization scheme is axiomatically and
logically the clearest, there is an inconvenience.  It is not well suited to
take into account the running of coupling strengths as functions of $q^2$
in consequence of renormalization group evolution.  This deficiency is
removed by the star $(\star)$ scheme of Kennedy and Lynn [17] which we
adopt for the rest of the lectures.  This preserves all the good features
of the on-shell scheme.  Yet, as will be clear below, it is able to write
$q^2$-dependent 1-loop physical amplitudes in terms of running coupling
strengths.

First, we return to the pi-functions and define them  for currents rather
than vector bosons.  We work in the $SU(2) \times U(1)$ theory and assume
weak isospin invariance.  Thus $\Pi^{AB} (q^2)$ is defined by
$$
i \Pi^{AB}_{\mu\nu} (q) \equiv \int d^4x e^{iq\cdot x} \langle \Omega
|J^A_\mu (x) J^B_\nu (0)|\Omega\rangle = i\Pi_{AB} (q^2) \eta_{\mu\nu} +
q_\mu q_\nu ~{\rm terms}.
\eqno (48)
$$
with (43) and (10) as well as the following figures, we have
$$
\begin{array}{r}
\Pi_{\gamma\gamma} = e^2 \Pi_{QQ},
{}~~~~~~~~~~~~~~~~~~~~~~~~~~~~~~~~~~~~~~~~ (49a) \\[2mm]
\Pi_{ZZ} = \displaystyle \frac {e^2} {c^2_\theta s^2_\theta} (\Pi_{33} -
2s^2_\theta \Pi_{3Q} + s^4_\theta \Pi_{QQ}),
{}~~~~~~~~~~~~~~~~~~~~~~~~~~~~~~~~~~~~~~~~ (49b) \\[2mm]
\Pi_{WW} = \displaystyle \frac {e^2} {2s^2_\theta} (\Pi_{11} + \Pi_{22}) =
\displaystyle \frac {e^2} {s^2_\theta} \Pi_{11},
{}~~~~~~~~~~~~~~~~~~~~~~~~~~~~~~~~~~~~~~~ (49c) \\[2mm]
\Pi_{\gamma Z} = \displaystyle \frac {e^2} {s_\theta c_\theta} (\Pi_{3Q} -
s^2_\theta \Pi_{QQ}). ~~~~~~~~~~~~~~~~~~~~~~~~~~~~~~~~~~~~~~~~ (49d)
\end{array}
$$
These are the unrenormalized $\Pi$-functions which are ultraviolet
divergent and hence regulator-dependent.  $EM$ gauge invariance dictates
that an on-shell photon is a pure state, i.e. $\Pi_{XQ} = 0 ~\forall~ X$.

Next, concentrate on $\Pi_{QQ} (q^2)$.  Since the photon self-energy has
no zero mass pole, $\Pi_{QQ} (q^2) \propto q^2$ as $q^2 \rightarrow 0$ and
it is convenient to define
$$
\Pi'_{QQ} (q^2) \equiv \displaystyle \frac {\Pi_{QQ} (q^2)} {q^2},
$$
where $\Pi'_{QQ} (0)$ is finite.  Now, from (39), the transverse part of
the complete (dressed) photon propagator acting between two physical lines
is
$$
e^2 (D^{\mu\nu}_{\gamma\gamma})_{tr} = - \displaystyle \frac
{i\eta^{\mu\nu}} {q^2 (1 - e^2 \Pi'_{QQ} (q^2))} \equiv - \displaystyle
\frac {i\eta^{\mu \nu}} {q^2} e^2_\star (q^2).
\eqno (50)
$$

\vspace* {3 cm}
\begin{center}
Fig. 18. Dressed photon propagator.
\end{center}

\noindent (50) introduces the running QED electric charge (or starred charge)
$$
e^2_\star (q^2) \equiv \displaystyle \frac {e^2} {1 - e^2 \Pi'_{QQ} (q^2)}
\simeq e^2 [1 + e^2 \Pi'_{QQ} (q^2)],
\eqno (51)
$$
the second step keeping only linear $\Pi$ terms to 1-loop.  The first RHS,
by the way, is the classic formula of Gell-Mann and Low.  Thus the value
of $\alpha_{EM}$, measured from $(g-2)_e$ or the A.C. Josephson effect is
$\alpha_{EM} \simeq \displaystyle {1 \over 4\pi} e^2_\star (0)$ whereas
that measured at LEP 1 on
the mass of the $Z$ corresponds to $\displaystyle \frac {1} {4\pi}
e^2_\star (M^2_Z)$.  In fact,
$$
\alpha^{-1}_{\star EM} (q^2) = \alpha^{-1}_{EM} - 4\pi [\Pi'_{QQ} (q^2) -
\Pi'_{QQ} (0)].
\eqno (52)
$$
The interesting point in (52) is that no explicit renormalization of the
pi-function is necessary since the particular combination is $UV$ finite.
As a specific application of (52), consider the 1-loop fermionic
contribution to $\alpha_\star (M^2_Z)$.  An explicit calculation yields
$$
\alpha^{-1}_{\star EM} (M^2_Z) - \alpha^{-1}_{EM} = - \displaystyle \frac
{1} {3\pi} \sum_f Q^2_f N^C_f \left(\ell n \displaystyle \frac {M^2_Z}
{M^2_f} - \frac {5}{3}\right).
\eqno (53)
$$
It is noteworthy that the RHS of (53) is the same as
$\delta\alpha_{EM}/\alpha^2_{EM}$ in the on-shell renormalization
scheme.  The evolution in (53), with all known fermions put in, brings
$\alpha^{-1}_{EM}$ from $\simeq 137.0$ down to $\alpha^{-1}_{\star EM}
(M^2_Z) \simeq 128.8$ and this is in brilliant agreement with experiment.
\bigskip

\noindent {\it Scattering}

\bigskip
\nobreak
Next, let us do a similar exercise for the neutral current coupling.  We
have done a detailed analysis for the $CC$ induced muon decay in the
on-shell scheme.  That is convenient there since all relevant energies are
low.  In contrast, physical neutral current scattering (like $CC$
scattering) is a highly $q^2$-dependent phenomenon and a treatment of its
renormalization in the star scheme is instructive.  The matrix element of
the $NC$ interaction between two physical scattering states can be
formally written to 1-loop as in Fig. 19.

\vspace* {4 cm}
\begin{center}
Fig. 19. $NC$ scattering with 1-loop oblique corrections.
\end{center}

We exclude Lorentz indices and spinor structures since we want to
obtain a compact form for the vacuum polarization insertions which only
modify the gauge boson propagators.  This is why they are called
``oblique'' as opposed to the ``direct'' vertex and box graph corrections.
It is also a fact that, except in the $Zb\bar b$ vertex,
these are the only significant 1-loop contributions in LEP physics.  Let us use
bare masses
$m^2_{\gamma b} = 0$, $m^2_{Wb} = e^2 v^2/(4s^2_\theta)$, $m^2_{Zb} = e^2
v^2/(4s^2_\theta c_\theta^2)$.  Thus we can write (including the QED part):
\begin{eqnarray}
M_{NC} &=& e^2 Q \displaystyle \frac {-i} {q^2} Q' + e^2 Q \displaystyle
\frac {-i} {q^2} i\Pi_{\gamma\gamma} (q^2) \displaystyle \frac {-i}
{q^2} Q' \nonumber \\[2mm]
& &
+ \displaystyle \frac {e^2} {c^2_\theta s^2_\theta} (T_3 -
s^2_\theta Q) \displaystyle \frac {-i} {q^2 - M^2_{Zb}} (T'_3 - s^2_\theta
Q') \nonumber \\[2mm]
& & + \displaystyle \frac {e^2} {c^2_\theta s^2_\theta} (T_3 - Qs^2_\theta)
\displaystyle \frac {-i} {q^2 - M^2_{Zb}} \Pi_{ZZ}(q^2) \displaystyle
\frac {-i}
{q^2 - M^2_{Zb}} (T'_3 - Q' s^2_\theta) \nonumber \\[2mm]
& & + \displaystyle \frac {e^2}
{c_\theta s_\theta} Q \displaystyle \frac {-i} {q^2} i\Pi_{\gamma Z} (q^2)
\displaystyle \frac {-i} {q^2 - M^2_{Zb}} (T'_3 - Q' s^2_\theta) \nonumber
\\[2mm] & & + \displaystyle \frac {e^2} {c_\theta s_\theta} (T_3 - Q
s^2_\theta) \displaystyle \frac {-i} {q^2 - M^2_{Zb}} i\Pi_{\gamma Z}
(q^2) \displaystyle \frac {-i} {q^2} Q' \nonumber
\end{eqnarray}

\newpage

\begin{eqnarray}
&=& e^2 QQ' \left(\displaystyle \frac {-i} {q^2}\right) \left[1 +
\displaystyle \frac {\Pi_{\gamma\gamma}} {q^2} (q^2)\right] +
\displaystyle \frac {e^2} {s^2_\theta c^2_\theta} (T_3 - s^2_\theta Q) \cdot
\nonumber \\[2mm]
& & (T'_3 - s^2_\theta Q') \left(\displaystyle \frac {-i} {q^2 - M^2_{Zb}}
\right) \left[1 + \displaystyle \frac {\Pi_{ZZ} (q^2)} {q^2 -
M^2_{Zb}}\right] \nonumber \\[2mm]
& & + \displaystyle \frac {e^2} {c_\theta s_\theta} \left[Q(T'_3 - Q'
s^2_\theta) + Q'(T_3 - Qs^2_\theta)\right] \left(\displaystyle \frac {-i}
{q^2 - M^2_{Zb}}\right) \Pi'_{\gamma Z} (q^2) \nonumber \\[2mm]
&\simeq& e^2_\star (q^2) Q\left(\displaystyle \frac {-i} {q^2}\right)Q' +
\displaystyle \frac {e^2} {s^2_\theta c^2_\theta} (T_3 - s^2_\theta Q)
\displaystyle \frac {-i} {q^2 - M^2_{Zb} - \Pi_{ZZ} (q^2)} \cdot \nonumber
\\[2mm] & & \left[T_3 -
Q's^2_\theta \left(1 - \displaystyle \frac {c_\theta} {s_\theta}
\Pi'_{\gamma Z} (q^2)\right)\right] \nonumber \\[2mm]
& & + \displaystyle \frac {e^2} {s^2_\theta c^2_\theta} \left[T_3 - Q
s^2_\theta \left(1 - \displaystyle \frac {c_\theta} {s_\theta}
\Pi'_{\gamma Z} (q^2)\right)\right] (T'_3 - s^2_\theta Q'). \nonumber
\end{eqnarray}

Now define $M^2(q^2) \equiv M^2_{Zb} + \Pi_{ZZ} (q^2)$.  At $q^2 = M^2_Z$
this becomes the on-mass shell renormalization scheme definition $M^2_Z =
M^2_{Zb} + \Pi_{ZZ} (M^2_Z)$.  Thus
$$
M^2 (q^2) = M^2_Z + \Pi_{ZZ} (q^2) - \Pi_{ZZ} (M^2_Z).
\eqno (54)
$$
Also define
$$
s^2_\star (q^2) \equiv  s^2_\theta \left(1 - {c_\theta \over s_\theta}
\Pi'_{\gamma Z} (q^2)\right),
\eqno (55)
$$
i.e.
$$
c^2_\star (q^2) \equiv c^2_\theta \left(1 - {s_\theta \over c_\theta}
\Pi'_{\gamma Z} (q^2)\right).
$$
Since we are working to linear terms in $\Pi$,
$$
\begin{array}{l}
(T_3 - s^2_\theta Q) \left[T'_3 - Q' s^2_\theta \left(1 -
\displaystyle{c_\theta
\over s_\theta} \Pi'_{\gamma Z} (q^2)\right)\right] \\[2mm]
{}~~~~~~~~~~~+ \left[T_3 -
Qs^2_\theta \left(1 - \displaystyle{c_\theta \over s_\theta} \Pi'_{\gamma Z}
(q^2)\right)\right] (T'_3 - Q' s^2_\theta) \\[2mm]
{}~~~~~~~~~~~\simeq \left[T_3 - Qs^2_\theta \left(1 - \displaystyle
{c_\theta \over s_\theta}
\Pi'_{\gamma Z} (q^2)\right)\right] \left[T'_3 - Q's^2_\theta \left(1 -
\displaystyle {c_\theta \over s_\theta} \Pi'_{\gamma Z}
(q^2)\right)\right] \\[2mm]
{}~~~~~~~~~~~= (T_3 -
Qs^2_\star (q^2)) (T'_3 - Q' s^{\star^2} (q^2)).
\end{array}
$$
One can then rewrite $M_{NC}$ as follows:
$$
M_{NC} = e^2_\star (q^2) Q\left(- {1 \over q^2}\right) Q' + {e^2 \over
c^2_\theta s^2_\theta} \left[T_3 - Qs^2_\star (q^2)\right] {-i \over q^2 -
M^2 (q^2)} [T'_3 - Q' s^2_\star (q^2)].
$$

We define the $Z$ wave function renormalization constant as the residue of
the pole at $q^2 = M^2_Z$ of $i$ times the $Z$-propagator, i.e.
$$
{1 \over q^2 - M^2_Z - \Pi_{ZZ} (q^2) + \Pi_{ZZ} (M^2_Z)} \equiv {Z_Z
\over q^2 - M^2_{Z^\star} (q^2)},
$$
where
$$
Z_Z \simeq 1 + {d \over dq^2} \Pi_{ZZ} (q^2)\Big|_{q^2 = M^2_Z}.
$$
Moreover, define a running wave function renormalization constant $Z_{Z
\star} (q^2)$ by
$$
{e^2_\star \over s^2_\star c^2_\star} Z_{Z^\star} \equiv {e^2 \over
s^2_\theta c^2_\theta} Z_Z.
\eqno (56)
$$
Calculating, we find
\begin{eqnarray}
Z_{Z^\star} (q^2) &=& Z_Z \left[1 - \pi'_{\gamma\gamma} (q^2) -
{c^2_\theta - s^2_\theta \over c_\theta s_\theta} \Pi'_{\gamma Z}
(q^2)\right]  \nonumber \\[2mm]
&\simeq& 1 + {d \over
dq^2} \Pi_{ZZ} (q^2)\Big|_{q^2 = M^2_Z} - \Pi'_{\gamma\gamma} (q^2) -
{c^2_\theta -s^2_\theta \over s_\theta c_\theta} \Pi'_{\gamma Z}
(q^2). \nonumber
\end{eqnarray}
This makes
$$
M_{NC} = e^2_\star Q \left(- {i \over q^2}\right) Q' + {e^2_\star \over
c^2_\star s^2_\star} (T_3 - Qs^2_\star) {Z_{Z^\star} \over q^2 -
M^2_{Z^\star}} (T'_3 - Q' s^2_\star)
\eqno (57)
$$
which is just the tree-level formula except that all bare quantities are
replaced by starred renormalized quantities.

\newpage

Exactly similar considerations can be followed for $CC$-induced
scattering.  Now one has the situation of Fig. 20.

\vspace* {5 cm}
\begin{center}
Fig. 20. $CC$ scattering with 1-loop oblique corrections.
\end{center}

\noindent Thus
$$
M_{CC} = {e^2_\star \over 2s^2_\star} I_+ {Z_{W^\star} \over q^2 -
M^2_{W^\star}} I_-,
\eqno (58)
$$
where
$$
{e^2_\star \over s^2_\star} Z_{W^\star} = {e^2 \over s^2_\theta} Z_W.
$$
Hence the renormalized version of (35) is
$$
A^{\mu\mu}_{FB} (M^2_Z) = {3[1-4s^2_\star (M^2_Z)]^2 \over [1 + \{1 -
4s^2_\star (M^2_Z)\}^2]^2}.
\eqno (59)
$$

We have to consider $M_{CC}$ and $M_{NC}$ at $q^2 = 0$.  After
incorporating 1-loop oblique corrections, one has an effective Lagrangian
density
$$
{\cal L}^{\rm Weak}_{EFF} = 4 {G_\mu \over \sqrt {2}} [J^+_{\nu L} J^{\nu
-}_L + \rho(J^3_{\nu L} - s^2_\star (0) J^Q_\nu) (J^{\nu 3}_L - s^2_\star
(0) J^{\nu Q})],
$$
with $\rho$ being the ratio of the nonelectrodynamic part of $M_{NC}$ to
the corresponding $M_{CC}$ at $q^2 = 0$.  Thus, with this ${\cal L}^{\rm
Weak}_{EFF}$ for instance,
$$
\Gamma_Z = \displaystyle {Z_{Z^\star} (M^2_Z) \alpha_\star (M^2_Z) \over
6s^2_\star (M^2_Z) c^2_\star (M^2_Z)} \sum_f [T_{3f} - s^2_\star (M^2_Z)
Q_f]^2 N^C_f,
$$
where, after QCD corrections,
\begin{eqnarray}
N^C_\ell &=& 1 + {3\alpha_{EM} (M_Z) \over 4\pi} Q^2_\ell, \nonumber
\\[2mm] N^C_q &=& \left[1 + {3\alpha_{EM} (M_Z) \over 4\pi} Q^2_q\right]
\left[1 - {\alpha_S (M_Z) \over \pi} + 0(\alpha^2_S)\right]. \nonumber
\end{eqnarray}

At $q^2 = 0$,
\begin{eqnarray}
q^2 - M^2_{Z^\star} &\rightarrow& -M^2_{Zb} - \Pi_{ZZ} (0) = - {e^2 \over
s^2_\theta c^2_\theta} \left[{v^2 \over 4} + \Pi_{33} (0)\right],\nonumber
\\[2mm] q^2 - M^2_{W^\star} &\rightarrow& - {e^2 \over 2s^2_\theta}
\left[{v^2 \over 4} + \Pi_{11} (0)\right]. \nonumber
\end{eqnarray}
Finally,
$$
\begin{array}{l}
M^{NQED}_{NC} (q^2 = 0) = [T_3 - s^2_\star (0) Q] \left[\displaystyle {v^2
\over 4} +
\Pi_{33} (0)\right]^{-1} [T_3 - s^2_\star (0) Q], \\[2mm]
M_{CC} (q^2 = 0) = {1\over2} T_+ \left[\displaystyle {v^2 \over 4} + \Pi_{11}
(0)\right]^{-1} T_-.
\end{array}
$$
Excluding the group theory factors,
$$
\rho = \left[ {v^2 \over 4} + \Pi_{11} (0)\right] \left[ {v^2 \over 4} +
\Pi_{33} (0)\right]^{-1} \simeq 1 + {4 \over v^2} \left[\Pi_{11} (0) -
\Pi_{33} (0)\right]
$$
to 1-loop.

We can now consider the $Zf\bar f$ vertex.  Since the outside
coupling is fixed to be $G_\mu$, as in (16), for the renormalized
couplings we need to take
$$
v_f \rightarrow \sqrt{\rho} [T_3 - Q_f s^2_\star (q^2)],
\eqno (60a)
$$
$$
a_f \rightarrow \sqrt{\rho} T_3,
\eqno (60b)
$$
$$
\rho = 1 + {4\pi \alpha_{EM} \over s^2_\theta c^2_\theta M^2_Z}
\left[\Pi_{11} (0) - \Pi_{33} (0)\right].
\eqno (60c)
$$
Now we need only specify $s^2_\star (q^2)$.

\noindent $sin^2 \theta_W$ \\
\bigskip

\nobreak
We discuss the ``renormalization'' of the sine of the Weinberg angle $\sin
\theta_W$.  The tree level $s_\theta$ is no longer an operative parameter
and we need a definition of $\sin \theta_W$ via a renormalized physical
process.  Veltman and Passarino like $\sin^2\theta_W \equiv 1 - M_W^2
M^{-2}_Z$, but the disadvantage there is that $M_W$ not well-measured as
yet.  Another approach is to define $\sin^2\theta_W$ as the ratio of
coupling constants renormalized by the $\overline{MS}$ scheme as in QCD,
but the problem here is that it cannot be simply related to physical
observables.  We want to generalize the tree-level relation (15) and {\it
define}
$$
\sin(2\theta_W)\big|_Z \equiv \left({4\pi \alpha_\star (M^2_Z) \over
\sqrt{2} G_\mu M^2_Z}\right)^{1/2},
\eqno (61)
$$
a definition which clearly relates $\sin^2\theta_W$ to physically
observable quantities.

One should first calculate $s^2_\star (q^2) - \sin^2\theta_W\big|_Z$.
Recall from (55) and (49d) that
$$
s^2_\star = {g^{\prime 2} \over g^2 + g^{\prime 2}} - e^2[\Pi'_{3Q} (q^2)
- s^2_\theta \Pi'_{QQ} (q^2)].
\eqno (62)
$$
Now (51) -- (53) can be rewritten as
$$
{\delta \alpha_{\star EM} (M^2_Z) \over \alpha_{EM}} \simeq e^2 \Pi'_{QQ}
(M^2_Z).
\eqno (63)
$$
Furthermore, the import of (44) is that -- in the oblique approximation --
$$
G_{\mu \star} (0) \simeq G_\mu \left(1 - {\Pi_{WW} (0) \over
M_{W^2}}\right).
$$
Thus
$$
{\delta G_{\mu \star} \over G_\mu} \simeq - {e^2 \over s^2_\theta}
\Pi_{11} (0).
\eqno (64)
$$
Again, from (36),
\begin{eqnarray}
M^2_Z &=& M^2_{Zb} \left[1 + {\Pi_{ZZ} (M^2_Z) \over M^2_{Zb}}\right]
\nonumber \\[2mm]
&=& {1\over4} (g^2 + g^{\prime 2})v^2 \left[1 + {e^2 \over s^2_\theta
c^2_\theta M^2_Z} \left\{\Pi_{33} (M^2_Z) - 2s^2_\theta \Pi_{3Q} (M^2_Z) +
s^4_\theta \Pi_{QQ} (M^2_Z)\right\}\right]. \nonumber
\end{eqnarray}
Thus
\begin{eqnarray}
\delta(\sin^2\theta_W) = 2\sin\theta_W\cos\theta_W \delta\theta_W &=&
{\sin\theta_W \cos\theta_W \delta(\sin2\theta_W) \over \cos^2\theta_W
- \sin^2\theta_W} \nonumber \\[2mm]
&=& {2s^2_\theta c^2_\theta \over c^2_\theta - s^2_\theta}
{\delta (\sin2\theta_W) \over \sin2\theta_W}. ~~~~~~~~~~~~~~~~~~~~~ (65)
\nonumber
\end{eqnarray}

Return to (15) and write it as
$$
\sin2\theta_{Wb} = \left({4\pi\alpha_{EMb} \over \sqrt{2} G_{\mu b}
M_{Zb}}\right)^{1/2},
\eqno (66)
$$
$$
\delta(\sin2\theta_{Wb}) = {1\over2} \left({4\pi\alpha_{EMb} \over
\sqrt{2} G_{\mu b} M^2_{Zb}}\right)^{1/2} \left[{\delta\alpha_{EM} \over
\alpha_{EM}} - {\delta G_\mu \over G_\mu} - {\delta M^2_Z \over
M^2_Z}\right].
\eqno (67)
$$
Now, by using (61) to (64), we can express $\sin^2\theta_W\big|_Z$ as
\begin{eqnarray}
\sin^2\theta_W\big|_Z &=& \sin^2\theta_{Wb} + \delta(\sin^2\theta_{Wb})
\nonumber \\[2mm]
&=& \sin^2\theta_{Wb} + {s^2_\theta c^2_\theta \over c^2_\theta -
s^2_\theta} \Bigg[e^2 \Pi'_{QQ} (0) + {e^2 \over s^2_\theta M^2_W}
\Pi_{11} (0) \nonumber \\[2mm]
& &  - {e^2 \over s^2_\theta c^2_\theta M^2_Z} \Pi_{33} (M^2_Z) - 2s_\theta
\Pi_{3Q} (M^2_Z) + s^4_\theta \Pi_{QQ} (M^2_Z)\Bigg]. ~~~~~~~ (68)
\nonumber
\end{eqnarray}
Use (62) and (67) to write
\begin{eqnarray}
s^2_\star (q^2) - \sin^2\theta_W\big|_Z &=& {e^2 \over c^2_\theta -
s^2_\theta} \Bigg[{\Pi_{33} (M^2_Z) - 2s^2_\theta \Pi_{3Q} (M^2_Z) -
\Pi_{11} (0) \over M^2_Z} \nonumber \\[2mm]
& & ~~~~~~~~~~~~- (c^2_\theta - s^2_\theta) \Pi'_{3Q}
(q^2)\Bigg] \nonumber \\[2mm]
& & + {e^2 s^2_\theta \over c^2_\theta - s^2_\theta} \Big[s^2_\theta
\Pi'_{QQ} (M^2_Z) - c^2_\theta \Pi'_{QQ} (0) \nonumber \\[2mm]
& & ~~~~~~~~~~~~+ (c^2_\theta - s^2_\theta)
\Pi'_{QQ} (q^2)\Big]. ~~~~~~~~~~~~~~~~~~~~ (69) \nonumber
\end{eqnarray}
\bigskip

\noindent {\it W-mass renormalization} \\
\bigskip

\nobreak
We have already seen that
$$
M^2_W = M^2_{Wb} + \Pi_{WW} (M^2_W) = M^2_{Wb} + {e^2 \over s^2_\theta}
\Pi_{11} (M^2_W).
\eqno (70)
$$
Furthermore,
$$
M^2_{Wb} = M^2_{Zb} \cos^2 \theta_{Wb}.
\eqno (71)
$$
Using (68), (70) and (71)
\begin{eqnarray}
M^2_W &=& M^2_Z \cos^2\theta_W\big|_Z - {e^2c^2_\theta \over s^2_\theta
(c^2_\theta - s^2_\theta)} \Bigg[\Pi_{33} (M^2_Z) - 2s^2_\theta \Pi_{3Q}
(M^2_Z) \nonumber \\[2mm]
& & ~~~~~~~~~~~~~~~~~~~~~~~~~~~~~~~~~~~- {s^2_\theta \over c^2_\theta}
\Pi_{11} (0) - {c^2_\theta
s^2_\theta \over c^2_\theta} \Pi_{11} (M^2_W)\Bigg]. ~~~~~ (72)
\nonumber
\end{eqnarray}
\bigskip

\begin{itemize}
\item {\bf Introduction to Oblique Parameters} \\
\end{itemize}
\smallskip

\nobreak
Within the framework of the obliqueness approximation, the three oblique
parametes [16] can be defined [18] as linear combinations of
$\Pi$-functions, defined at $q^2 = 0$ and $q^2 = M^2_Z$.  We first
introduce the hypercharge current $J_\mu^Y$ as a linear combination of the
electromagnetic current and the third weak isospin current:
$$
J^Q_\mu = J^3_\mu + {1\over2} J^Y_\mu.
\eqno (73)
$$
Now we define
\begin{eqnarray}
S &\equiv& {16\pi \over M^2_Z} \left[\Pi_{33} (M^2_Z) - \Pi_{33} (0) -
\Pi_{3Q} (M^2_Z)\right] \nonumber \\[2mm]
&=& {8\pi \over M^2_Z} \left[\Pi_{3Y} (0) - \Pi_{3Y} (M^2_Z)\right],
{}~~~~~~~~~~~~~~~~~~~~~~~~~~~~~~~~~~~~~ ~~~~~~~~~~ (74a) \nonumber
\\[2mm]
T &\equiv& {4\pi \over s^2_\theta c^2_\theta} M^{-2}_Z [\Pi_{11} (0) -
\Pi_{33} (0)], ~~~~~~~~~~~~~~~~~~~~~~~~~~~~~~~~~~~~~~~~~~~~~~~
(74b) \nonumber \\[2mm]
U &\equiv& {16\pi \over M^2_W} \left[\Pi_{11} (M^2_W) - \Pi_{11}
(0)\right] - {16\pi \over M^2_Z} [\Pi_{33} (M^2_Z) - \Pi_{33} (0)].
{}~~~~~~~~~~~~~~~ (74c) \nonumber
\end{eqnarray}

There are other definitions [19] of these parameters.  Particularly
popular is [20] the set $\epsilon_1,\epsilon_2,\epsilon_3$ where
$$
\epsilon_1 = \alpha_{EM} T,
\eqno (75a)
$$
$$
\epsilon_2 = - {\alpha_{EM} \over 4s^2_\theta} U,
\eqno (75b)
$$
$$
\epsilon_3 = {\alpha_{EM} \over 4s^2_\theta} S.
\eqno (75c)
$$
More general definitions of these parameters, going outside the
obliqueness approximation, also exist [21].

1. There are two important aspects of the oblique parameters which should
be highlighted.  $T$ and $U$ receive nonzero contributions from the
violation of weak isospin and are finite on account of the weak isospin
symmetric nature of the divergence terms.  $S$ originates from the mixing
between the weak hypercharge and the third component of weak isospin as a
consequence of the spontaneous symmetry breakdown mechanism.  Only soft
operators (i.e. those with scale dimensionality less than four) are
involved in the latter process.  By Symanzik's theorem [22], these do not
contribute to the leading divergences and therefore $S$ is free of them.
Furthermore, the nonleading divergences cancel out in the difference
between $\Pi_{3Y} (M^2_Z)$ and $\Pi_{3Y} (0)$ leaving a finite $S$.

2. The LHS of (48) can be rewritten by inserting a complete set of states as
$$
(2\pi)^4 \sum_n \delta^{(4)} (q - p_n) \langle \Omega|J^A_\mu (0)|n\rangle
{}~\langle n|J^B_\nu (0)|\Omega\rangle.
$$
Any new physics effect from beyond $SM$ would come from a new set of
states and hence would be linearly adding to that from $SM$, i.e. the
$\Pi_{AB}$ functions receive contributions from different sources
additively.  This enables one to define $\tilde \Pi_{AB} = \Pi_{AB} -
\Pi^{SM}_{AB}$.  Of course, $\Pi^{SM}_{AB}$ depends on the yet unknown top
and Higgs masses quadratically and logarithmically in respective order,
the latter being a consequence of Veltman's screening theorem [23].

The relationship between the oblique parameters and observables can be
obtained by rewriting the $\Pi$-functions in terms of $S,T$ and $U$ in
Eqs. (60) -- (72).  Specifically, (60c) changes to
$$
\rho = 1 + \alpha_{EM} T.
\eqno (76)
$$
Furthermore, (69) changes to (with $q^2 = M^2_Z$)
$$
s^2_\star (M^2_Z) - \sin^2\theta_W\big|_Z = {\alpha_{EM} \over c^2_\theta
- s^2_\theta} \left({1\over4} S - s^2_\theta c^2_\theta T\right)
\eqno (77)
$$
and (72) changes to
$$
M^2_W = M^2_Z \cos^2\theta_W\big|_Z + {\alpha_{EM} M^2_Z c^2_\theta \over
c^2_\theta - s^2_\theta} \left(-{1\over2} S + c^2_\theta T + {c^2_\theta -
s^2_\theta \over 4s^2_\theta} U\right).
\eqno (78)
$$

Moreover, one can split
$$
(S,T,U) = (S,T,U)^{SM} + (\tilde S,\tilde T,\tilde U)
$$
and consequently rewrite (76) -- (78) as
$$
\rho = \rho^{SM} + \alpha \tilde T,
\eqno (79)
$$
$$
s^2_\star (M^2_Z) = [s^2_\theta (M^2_Z)]^{SM} + {\alpha_{EM} \over
4(c^2_\theta - s^2_\theta)} (\tilde S - 4c^2_\theta s^2_\theta \tilde T),
\eqno (80)
$$
$$
M^2_W = (M^2_W)^{SM} + {\alpha_{EM} M^2_Z c^2_\theta \over 4s^2_\theta
(c^2_\theta - s^2_\theta)} \left[4c^2_\theta s^2_\theta \tilde T -
2s^2_\theta \tilde S + (c^2_\theta - s^2_\theta)\tilde U\right].
\eqno (81)
$$
Here $\rho^{SM}$, $[s^2_\star (M^2_Z)]^{SM}$ and $(M^2_W)^{SM}$ are these
quantities, calculated to 1-loop in the standard model in terms of
$\alpha_{EM}$, $G_\mu$ and $M_Z$ as well as fermionic and Higgs masses.

The determination of $\tilde S$ and $\tilde T$ from experiment is best
done as follows.  One calculates the differential cross section for the
process $e^+e^- \rightarrow f\bar f$ in the $Z$ lineshape region with the
$\star$-scheme effective Lagrangian and the couplings $v_f$, $a_f$ of (60
$a,b$).  The latter are rewritten in terms of $\tilde S$ and $\tilde T$
which are obtained by detailed fits with the millions of accumulated data
points.  The latest fit with 5 million data points yields [24], for $m_t =
160$ GeV and $m_H = 100$ GeV,
$$
\begin{array}{l}
\tilde S = -0.49 \pm 0.31 \\[2mm]
\tilde T = -0.10 \pm 0.32.
\end{array}
\eqno (82)
$$
Thus $\tilde T$ is compatible with the null value of $SM$ whereas in
$\tilde S$ there is hint of a nonzero value at the 1.5~$\sigma$ level.  The
variations of these numbers with changes in $m_t$ and $m_H$ have also been
studied [24].  In particular $\tilde S$ is insensitive to variations in
$m_t$ in the range of interests.
The extraction of $\tilde U$ is rather inaccurate because
of its sensitive dependence [vide (81)] on the $W$-mass which is rather
poorly known.  Using $M_W = 80.24 \pm 0.10$ GeV and the $\tilde S,\tilde
T$ values of (82), one is led to
$$
\tilde U = -0.11 \pm 0.82.
\eqno (83)
$$
The error will be significantly reduced once the $W$-mass is better known.

The oblique parameters are powerful probes for certain types of new
physics.  Any scenario which goes beyond $SM$ will have particles heavier
than those in the latter.  A most important question [25] is how the effects of
such particles on $SM$ processes would act as their masses are made larger
and larger.  If these mass terms are $SU(2)_L \times U(1)_Y$
gauge-invariant, those effects decouple as inverse powers of the heavy
masses and the new physics is of a decoupling type.  An example of this is
the supersymmetric extension of the $SM$.  Contrariwise, for mass terms
that are gauge-variant vis-a-vis $SU(2)_L \times U(1)$ transformations,
those effects do not decouple even as the heavy masses become larger.
This type of nondecoupling new physics is caused by extra heavy chiral
fermion generations or condensate models such as technicolor.

In general, a decoupling type of new physics leads to rather small
(compared to unity) values of $\tilde S,\tilde T,\tilde U$ for the new
mass-scale in the sub-TeV to TeV region.  Thus knowledge of the oblique
parameters (with the kind of accuracy that is realistically feasible)
cannot significantly test or constrain such models.  This is, however, not
the case with nondecoupling new scenarios.  In particular, $\tilde S$ is a
rather sensitive probe for chiral fermion condensate models.
Specifically, most technicolor and extended technicolor scenarios predict
[26] large positive $S \geq 0.4$ and are disfavored by the data.
\bigskip

\begin{itemize}
\item {\bf Acknowledgements} \\
\end{itemize}
\smallskip

\nobreak
I would like to thank Jogesh Pati for inviting me to lecture in this
school and Liviana Forza for gently persuading me to write up these notes
in time for the proceedings.

\newpage

\begin{itemize}
\item References
\end{itemize}
\smallskip

\begin{enumerate}
\item[{[1]}] S.L. Glashow, Nucl. Phys. B\underbar{22} (1961) 579.  S.
Weinberg, Phys. Rev. Lett. \underbar{19} (1967) 1264.  A. Salam in {\it
Elementary Particle Theory}, ed. N. Svartholm (Stockholm, 1968), p367.
\item[{[2]}] H. Fritzsch, M. Gell-Mann, H. Lleutwyler, Phys. Lett.
\underbar{47} (1973) 365.  D. Gross and F. Wilczek, Phys. Rev. Lett.
\underbar{30} (1973) 1343.  H.D. Politzer, Phys. Rev. Lett. \underbar{30}
(1973) 1346.  S. Weinberg, Phys. Rev. Lett. \underbar{31} (1973) 494.
\item[{[3]}] P.W. Higgs, Phys. Lett. \underbar{12} (1964) 132; Phys. Rev.
Lett. \underbar{13} (1964) 508; Phys. Rev. \underbar{145} (1966) 1156.  F.
Englert and R. Brout, Phys. Rev. Lett. \underbar{13} (1964) 321.  G.S.
Guralnik, C.R. Hagen and T.W.B. Kibble, Phys. REv. Lett. \underbar{13}
(1964) 585.  T.W.B. Kibble, Phys. Rev. \underbar{155} (1967) 1554.
\item[{[4]}] Particle Data Group.  M. Aguilar-Benitez et al, Phys. Rev.
\underbar{D45} (1992), Part 2 (June 1992).
\item[{[5]}] L. Susskind, Phys. Rev. \underbar{D20} (1979) 2619.  S.
Weinberg, Phys. Rev. \underbar{D19} (1979) 1277.  S. Dimopoulos and L.
Susskind, Nucl. Phys. \underbar{B155} (1979) 237.  E. Eichten and K.D.
Lane, Phys. Lett. \underbar{90B} (1980) 125.  E. Farhi and L. Susskind,
Phys. Rep. \underbar{74} (1981) 277.
\item[{[6]}] P. Sikivie, L. Susskind, M. Voloshin and V. Zakharov, Nucl.
Phys. \underbar{B173} (1980) 189.
\item[{[7]}] N. Cabibbo, Phys. Lett. \underbar{10} (1963).  M. Kobayashi
and K. Maskawa, Prog. Theor. Phys. \underbar{49} (1973) 652.
\item[{[8]}] G. Coignet, talk given at the {\it XVI Int. Symp.
Lepton-Photon Interactions}, Cornell University, Ithaca, August 1993.
\item[{[9]}] Z. Hioki, Mod. Phys. Lett. \underbar{A7} (1992) 1009.
\item[{[10]}] {\it Z Physics at LEP 1} (ed. G. Altarelli, R. Kleiss and C.
Verzegnassi), CERN Report No. 89-08, Vols. 1,2 and 3.
\item[{[11]}] T. Kinoshita, {\it Proc. XIX Int. Conf. High Energy
Physics}, Tokyo (1978, ed. S. Homma et al).
\item[{[12]}] F. Jegerlehner, ``Physics of Precision Experiments with
Z's'' in {\it Progress in Particle and Nuclear Physics} (ed. A. F\"assler,
Pergamon).  W.F.L. Hollik, Fortschr. Phys. \underbar{38} (1990) 165.
\item[{[13]}] J. Fleisher and F. Jegerlehner, Phys. Rev. \underbar{D23}
(1981) 2001.  F. Jegerlehner in {\it Radiative Corrections in $SU(2)_L
\times U(1)$}, eds. B.W. Lynn and J.F. Wheater (World Scientific, 1984).
\item[{[14]}] Z. Hioki, Ref. 9.  G.
Rajasekaran in {\it Phenomenology of the Standard Model and Beyond} (eds.
D.P. Roy and Probir Roy, World Scientific, 1989), p407.
\item[{[15]}] G. Altarelli in {\it Phenomenology of the Standard Model and
Beyond, op. cit}, p383.
\item[{[16]}] B.W. Lynn, M.E. Peskin and R.G. Stuart in Physics at LEP,
LEP Jamboree, Geneva 1985 (ed. J. Ellis and R. Peccei, CERN Report No.
86-02, 1986).  M.E. Peskin and T.E. Takeuchi, Phys. Rev. Lett.
\underbar{65} (1990) 2963; Phys. Rev. \underbar{D46} (1992) 381.
\item[{[17]}] D.C. Kennedy and B.W. Lynn, Nucl. Phys. \underbar{B322}
(1989) 1.
\item[{[18]}] G. Bhattacharyya, S. Banerjee and P. Roy, Phys. Rev.
\underbar{D45} (1992) R729.
\item[{[19]}] D.C. Kennedy and P. Langacker, Phys. Rev. Lett.
\underbar{65} (1990) 2967.  W.J. Marciano and J.L. Rosner, Phys. Rev.
Lett. \underbar{65} (1990) 2963.
\item[{[20]}] G. Altarelli and R. Barbieri, Phys. Lett. \underbar{B253}
(1991) 161.  G. Altarelli, R. Barbieri and S. Jadach, Nucl. Phys.
\underbar{B369} (1992) 3.
\item[{[21]}] P. Roy in Proc. X DAE High Energy Physics Symposium, Pramana
J. Phys. (Suppl.), in press.
\item[{[22]}] K. Symanzik, Carg\'ese Lectures, Vol. 5 (1972).
\item[{[23]}] M. Veltman, Act. Phys. Polon. \underbar{B8} (1977) 475.
\item[{[24]}] M. Swartz, talk given at the XVI Int. Symp. Lepton-Photon
Interactions, Cornell University, Ithaca, August 1993.
\item[{[25]}] T.P. Cheng and L.F. Li, Phys. Rev. \underbar{D44} (1991)
1502.
\item[{[26]}] P. Langacker, University of Pennsylvania Report No.
UPR-0512-T, unpublished.  R. Sundrum and H.D. Hsu, Nucl. Phys.
\underbar{B391} (1993) 127.
\end{enumerate}
\end{document}